\documentclass[reprint,twocolumn,english,aps,prb,floatfix,groupedaddress,showpacs,superscriptaddress,citeautoscript,amsmath,amssymb]{revtex4-1}
\usepackage{amsmath}
\usepackage{amsbsy}
\usepackage{graphicx}
\usepackage{units}
\usepackage{xcolor} 
\usepackage{soul} 
\usepackage{amsmath}
\usepackage{amsfonts}
\usepackage{mathrsfs}
\usepackage{amsbsy}
\usepackage{bm}
\usepackage{babel}

\newcommand{\hh}{\ensuremath{H}}
\newcommand{\kk}{\ensuremath{K}}

\begin{document}

\title{Spin dynamics near a putative antiferromagnetic quantum critical point in Cu substituted BaFe$_2$As$_2$ and its relation to high-temperature superconductivity}

\author{M.~G.~Kim}\thanks{mgkim@lbl.gov}
\affiliation{Materials Sciences Division, Lawrence Berkeley National Laboratory, Berkeley, CA 94720, USA}
\author{M. Wang}
\affiliation{Department of Physics, University of California, Berkeley, CA 94720, USA}
\author{G. S. Tucker}
\affiliation{Ames Laboratory and Department of Physics and Astronomy, Iowa State University, Ames, IA, 50011, USA}
\author{P.~N.~Valdivia}
\affiliation{\mbox{Department~of~Materials~Science~and~Engineering,~University~of~California,~Berkeley,~CA~94720,~USA}}\author{D. L. Abernathy}
\affiliation{Quantum Condensed Matter Division, Oak Ridge National Laboratory, Oak Ridge, TN 37831, USA}
\author{Songxue~Chi}
\affiliation{Quantum Condensed Matter Division, Oak Ridge National Laboratory, Oak Ridge, TN 37831, USA}
\author{A.~D.~Christianson}
\affiliation{Quantum Condensed Matter Division, Oak Ridge National Laboratory, Oak Ridge, TN 37831, USA}
\author{A.~A.~Aczel}
\affiliation{Quantum Condensed Matter Division, Oak Ridge National Laboratory, Oak Ridge, TN 37831, USA}
\author{T.~Hong}
\affiliation{Quantum Condensed Matter Division, Oak Ridge National Laboratory, Oak Ridge, TN 37831, USA}
\author{T.~W.~Heitmann}
\affiliation{The Missouri Research Reactor, University of Missouri, Columbia, MO 65211, USA}
\author{S.~Ran}
\affiliation{Ames Laboratory and Department of Physics and Astronomy, Iowa State University, Ames, IA, 50011, USA}
\author{P.~C.~Canfield}
\affiliation{Ames Laboratory and Department of Physics and Astronomy, Iowa State University, Ames, IA, 50011, USA}
\author{E.~D.~Bourret-Courchesne}
\affiliation{Materials Sciences Division, Lawrence Berkeley National Laboratory, Berkeley, CA 94720, USA}
\author{A.~Kreyssig}
\affiliation{Ames Laboratory and Department of Physics and Astronomy, Iowa State University, Ames, IA, 50011, USA}
\author{D.~H.~Lee}
\affiliation{Materials Sciences Division, Lawrence Berkeley National Laboratory, Berkeley, CA 94720, USA}
\affiliation{Department of Physics, University of California, Berkeley, CA 94720, USA}
\author{A.~I.~Goldman}
\affiliation{Ames Laboratory and Department of Physics and Astronomy, Iowa State University, Ames, IA, 50011, USA}
\author{R.~J.~McQueeney}
\affiliation{Ames Laboratory and Department of Physics and Astronomy, Iowa State University, Ames, IA, 50011, USA}
\author{R.~J.~Birgeneau}
\affiliation{Materials Sciences Division, Lawrence Berkeley National Laboratory, Berkeley, CA 94720, USA}
\affiliation{Department of Physics, University of California, Berkeley, CA 94720, USA}
\affiliation{\mbox{Department~of~Materials~Science~and~Engineering,~University~of~California,~Berkeley,~CA~94720,~USA}}

\date{\today}


\begin{abstract}
We present the results of elastic and inelastic neutron scattering measurements on non-superconducting Ba(Fe${_{0.957}}$Cu${_{0.043}}$)${_2}$As${_2}$, a composition close to a quantum critical point between AFM ordered and paramagnetic phases. By comparing these results with the spin fluctuations in the low Cu composition as well as the parent compound BaFe$_2$As$_2$ and superconducting Ba(Fe$_{1-x}$Ni$_x$)$_2$As$_2$ compounds, we demonstrate that paramagnon-like spin fluctuations are evident in the antiferromagnetically ordered state of Ba(Fe$_{0.957}$Cu$_{0.043}$)$_2$As$_2$, which is distinct from the AFM-like spin fluctuations in the superconducting compounds. 
Our observations suggest that Cu substitution decouples the interaction between quasiparticles and the spin fluctuations. We also show that the spin-spin correlation length, ${\xi(T)}$, increases rapidly as the temperature is lowered and find ${\omega/T}$ scaling behavior, the hallmark of quantum criticality, at an antiferromagnetic quantum critical point.
\end{abstract}

\maketitle

\section{Introduction}

Inelastic neutron scattering (INS) measurements on the iron-arsenide parent BaFe$_2$As$_2$ compound show strong antiferromagnetic (AFM) spin fluctuations at temperature below the AFM ordering temperatures ($T_\mathrm{N}$) together with the evolution to paramagnetic fluctuations for $T>T_\mathrm{N}$.\cite{Harriger12} In the superconducting (SC) iron arsenides, such as Ba(Fe$_{1-x}$Ni$_x$)$_2$As$_2$, strong spin fluctuations exist, whether the system orders magnetically or not. These fluctuations look very similar to those observed in the AFM ordered parent compound, except for the onset of a superconducting spin resonance that appears below the superconducting transition temperature ($T_\mathrm{c}$).\cite{Liu12, Luo13} The observation of strong AFM spin fluctuations in the superconducting compounds invigorates the idea that the spin fluctuations may provide the pairing interaction for the Cooper pairing of quasiparticles\cite{Monthoux07,Wang11,Scalapino12,Hosono15}. 

As important as strong spin fluctuations seem to be, superconductivity emerges in the iron arsenides only if the AFM order is sufficiently suppressed to lower temperatures by means of external parameters, such as an elemental substitution.\cite{Johnston10,canfield_feas_2010, Dai15} For example, in Ba(Fe$_{1-x}$$TM_x$)$_2$As$_2$ with $TM$= Co or Ni, $T_\mathrm{N}$ is lowered and $T_\mathrm{c}$ rises with increasing substitution level. In earlier studies, beyond some threshold, $T_\mathrm{N}$ seemed to decrease below $T_\mathrm{c}$ down to zero temperature.\cite{Johnston10,canfield_feas_2010, Dai15,canfield09, ni08, li_superconductivity_2009, ni10} As $T_\mathrm{N} \rightarrow 0$, a zero temperature instability may exist between the AFM and paramagnetic states. In other words, an antiferromagnetic quantum critical point (AFM QCP) is anticipated. However, several neutron experiments show that the $T_\mathrm{N}$ is arrested by the appearance of superconductivity. Some studies have shown that the magnetism becomes paramagnetic again below $T_\mathrm{c}$ in higher Co substituted compounds, resulting in the back-bending of the $T_\mathrm{N}$ phase line.\cite{Fernandes10, Dai15} Other studies have demonstrated the discontinuous suppression of $T_\mathrm{N}$ at a non-zero temperature, implying an avoided AFM QCP in Ni and P substituted BaFe$_2$As$_2$ compounds.\cite{Lu13,Lu14, Hu15} However, interestingly, previous transport measurements\cite{Kasahara10}, NMR measurements\cite{Nakai10}, and penetration depth measurements\cite{Hashimoto_2012} in P substituted BaFe$_2$As$_2$ compounds show non-Fermi liquid behavior, pointing to the possible existence of a QCP. Such non-Fermi liquid behavior is one of the characteristics of the well-known heavy-fermion superconductors. In these materials, inelastic neutron scattering measurements found that the spin fluctuations show a singular behavior as well as ${\omega/T}$ scaling behavior, the hallmark of quantum criticality at the AFM QCP.\cite{Lohneysen_2007, Sachdev_book}

In the case of Cu substitution, the phase diagram of Ba(Fe$_{1-x}$Cu$_x$)$_2$As$_2$ looks very similar to that of Co or Ni substituted BaFe$_2$As$_2$ compounds.~\cite{canfield09, ni08, li_superconductivity_2009,ni10}  Ba(Fe$_{1-x}$Cu$_x$)$_2$As$_2$ exhibits the same kind of structural and magnetic transitions with $T_\mathrm{S} > T_\mathrm{N}$ as in other superconducting compounds.\cite{canfield09, ni10,Kim12} However, superconductivity is not observed down to the lowest measured temperature of 2K for any degree of Cu substitution. Therefore, unlike in the superconducting compounds, the AFM order is not hindered by occurrence of superconductivity and a putative AFM quantum critical point can be reached in Ba(Fe$_{1-x}$Cu$_x$)$_2$As$_2$, especially with 0.044 $< x \leq$ 0.047 as we show later in the results section. We can then investigate the consequences of the quantum criticality on the spin dynamics without an intervening superconducting state and seek possible connections between the spin fluctuations and superconductivity in Ba(Fe$_{1-x}$$TM_x$)$_2$As$_2$.

In this paper, we present a detailed study of the spin fluctuations over a wide range of temperature and energy transfers in the non-superconducting Ba(Fe${_{0.957}}$Cu${_{0.043}}$)${_2}$As${_2}$ compound. We explore, in particular, any potential connection to unconventional superconductivity in the iron arsenides. First, neutron diffraction is used to establish a phase diagram of Ba(Fe$_{1-x}$Cu$_x$)$_2$As$_2$ compounds. Then we use the inelastic neutron scattering technique and show that while the spin fluctuation spectra in Ba(Fe${_{0.957}}$Cu${_{0.043}}$)${_2}$As${_2}$ look similar to those in related superconducting derivatives for $T > T_\mathrm{c}$, the momentum-integrated local susceptibility, ${\chi ''(\omega)}$, exhibits paramagnetic fluctuations at all temperatures, which is distinct from the AFM-like spin fluctuations in the SC compounds. From a detailed study of the spin fluctuations over a wide range of temperatures, we also show that the spin-spin correlation length, ${\xi(T)}$, increases rapidly as we lower the temperature and discover ${\omega/T}$ scaling behavior over a wide range of temperatures, supporting the existence of a putative AFM QCP at 0.044 $< x \leq$ 0.047.

\section{Experiment}

Single crystals of  Ba(Fe$_{1-x}$Cu$_{x}$)$_{2}$As$_{2}$ ($x$ = 0.028, 0.039, 0.043, 0.044, and 0.047) were grown out of a FeAs self-flux using conventional high-temperature solution growth.\cite{ni10} Elemental analysis was performed at approximately 10 positions on each sample using wavelength dispersive spectroscopy providing a relative uncertainty of less than 5\%. For the determination of the phase diagram (shown in Fig.~\ref{PD}), single pieces of crystal were used in neutron measurements for $x$ = 0.039, 0.044, and 0.047 (a typical mass of approximately 100 mg) and co-aligned crystals were used for $x$ = 0.028 (2 crystals, total mass of 1.5 g) and 0.043 (9 crystals, total mass of 1.14 g). 

The neutron diffraction measurements were performed on the TRIAX triple-axis spectrometer at the University of Missouri Research Reactor and the HB3 triple-axis spectrometer at the High-Flux Isotope Reactor at the Oak Ridge National Laboratory. Samples with $x$ = 0.039, 0.044, and 0.047 were measured on the TRIAX. The beam collimators before the monochromator, between the monochromator and sample, between the sample and analyzer, and between the analyzer and detector were $60'-40'-40'-80'$, respectively. We used fixed $E_\mathrm{i} = E_\mathrm{f} = 14.7$ meV and two pyrolytic graphite filters to eliminate higher harmonics in the incident beam. For the $x$ = 0.028 and 0.043 samples, we used the HB3 spectrometer with $48'-60'-80'-120'$ collimation, fixed $E_\mathrm{i} = E_\mathrm{f} = 14.7$ meV, and 2 PG filters before the analyzer. 

Inelastic neutron scattering measurements on the $x$ = 0.028 and 0.043 samples were performed on the cold triple-axis spectrometer (CTAX), the HB3 triple-axis spectrometer at the High-Flux Isotope Reactor at the Oak Ridge National Laboratory, and the Wide Angular-Range Chopper Spectrometer (ARCS)\cite{Abernathy12} at the Spallation Neutron Source at Oak Ridge National Laboratory. At CTAX, we used a fixed $E_\mathrm{f} = 5$ meV with open$-$open$-$open$-$open collimation and a nitrogen-cooled Be filter was employed to eliminate higher harmonics. The experimental setup at HB3 was identical to that for diffraction measurements except that we used a fixed $E_\mathrm{f} = 14.7$ meV. For the time-of-flight measurement on ARCS, we used $E_\mathrm{i}=$ 50, 80, and 250 meV with $k_i$ parallel to the $\bm{c}$-axis. The incident Fermi chopper frequencies were 120 Hz and 420 Hz for $E_\mathrm{i} = 50$ and 80 meV and 360 Hz and 120 Hz for $E_\mathrm{i} = 250$ meV.

The samples were aligned such that the ($H$,~0,~$L$) reciprocal lattice plane was coincident with the scattering plane of the spectrometer. Measurements were performed using closed-cycle refrigerators between room temperature and the base temperature, $T = 5$ K of the refrigerator. All samples exhibited small mosaicities [$<0.4^{\circ}$ full-width-at-half-maximum (FWHM) measured at CTAX, and $<0.6^{\circ}$ FWHM measured at HB3] measured by rocking scans through the (4,~0,~0) nuclear peak, demonstrating high sample quality. In our measurements we define $\bm{Q} = \left(\hh,\kk,L\right) = \frac{2\pi}{a}\hh\hat{\imath} + \frac{2\pi}{a}\kk\hat{\jmath} + \frac{2\pi}{c}L\hat{k}$ where the orthorhombic lattice constants are $a \geq b \approx 5.6 $\,\AA~and $c\approx 13 $\,\AA. Note that Ba(Fe${_{0.972}}$Cu${_{0.028}}$)${_2}$As${_2}$ (`Cu028', $T_\mathrm{S} \approx 73$ K, $T_\mathrm{N} \approx 64$ K, no SC) compound studied in the current paper is the co-aligned set already used in Ref.~\onlinecite{Kim12}. We used it for additional measurements that were not presented in the previous report to compare with Ba(Fe${_{0.972}}$Cu${_{0.043}}$)${_2}$As${_2}$ (`Cu043') compound.

\section{Results}

\subsection{Phase Diagram}

\begin{figure}[t!]
\centering
\includegraphics[width=0.93\linewidth]{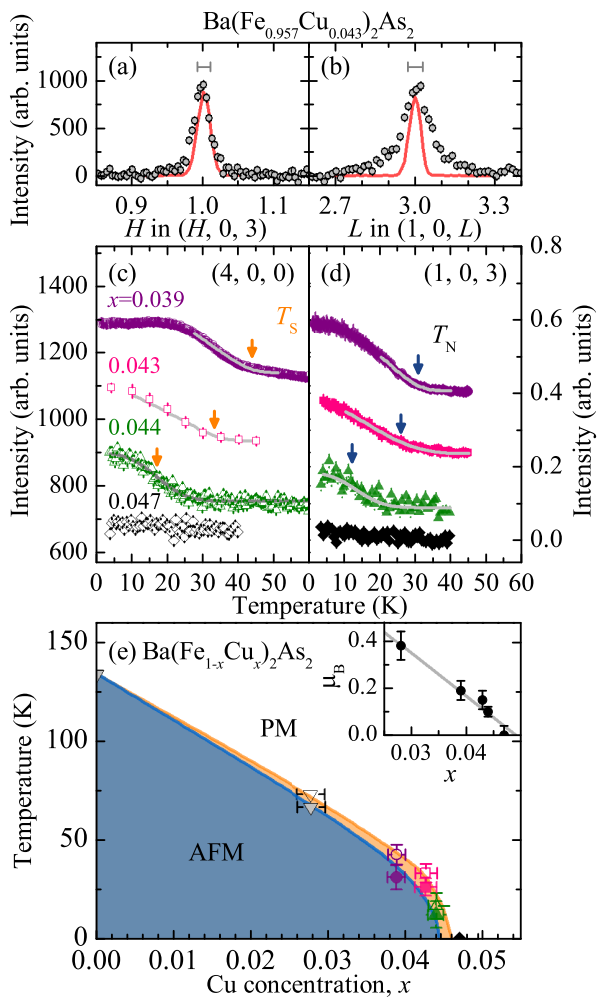} 
\caption{(a) and (b), The AFM peak in Ba(Fe$_{1-x}$Cu$_{x}$)$_2$As$_2$ with $x=0.043$, scanned through the $\bm{Q}_\mathrm{AFM}=(1,~0,~3)$ along the orthorhombic $\bm{a}$ and $\bm{c}$ directions, respectively. The lines present the same scans for $x = 0.028$. The bars indicate the instrument resolution. (c) Temperature dependent intensities of the (4,~0,~0) Bragg peak for $x =$ 0.039, 0.043, 0.044, and 0.047. (d) The AFM order parameter measured at $\bm{Q}_\mathrm{AFM}=(1,~0,~3)$ for $x =$ 0.039, 0.043, 0.044, and 0.047. The lines in (c) and (d) are the results of fits as described in the text. Arrows in (c) and (d) indicate $T_\mathrm{S}$ and $T_\mathrm{N}$, respectively. (e) The phase diagram of Ba(Fe$_{1-x}$Cu$_{x}$)$_2$As$_2$. The inset shows the ordered moments at $T = 0$. The error bars indicate the statistical errors of one standard deviation.}
\label{PD}
\end{figure}

Figures~\ref{PD}a and~\ref{PD}b show [$H$, 0, 0] and [0, 0, $L$] scans through the (1, 0, 3) magnetic Bragg peak for $x =$ 0.043 (circles) and $x =$ 0.028 (lines) measured at the HB3 instrument at the High-Flux Isotope Reactor at the Oak Ridge National Laboratory (ORNL). We find that the AFM ordering in $x =$ 0.043 is consistent with the spin-density wave order observed in the parent compound. While the line shape broadens along the orthorhombic $\bm{b}$ direction for $x \geq$ 0.039\cite{Kim12}, the peak width along the orthorhombic $\bm{a}$ direction is resolution-limited and comparable to that in both the $x = 0.028$ and 0.043 samples. However, the peak width along the $\bm{c}$ direction for the $x =$ 0.043 sample becomes 3 times broader than that for the $x =$ 0.028 sample. It has been proposed that the broadening along the $\bm{b}$ direction and the absence of incommensurate AFM order in Cu substituted compounds arise from disorder by Cu substitution which introduces spectral broadening of the Fermi-surfaces\cite{Kim12}. The broadening along the $\bm{c}$ direction supports this scenario. However, the resolution-limited peak width along the $\bm{a}$ direction is not consistent with the proposal and suggests that the Fermi-surface spectra along the $\bm{a}$ direction may be protected from broadening in this proposal. Further study is necessary to understand the relation between the disorder effect and the AFM ordering in iron arsenides. 

We show the structural and AFM order parameters for samples with $ 0.039 \leq x \leq 0.047$ in Figs.~\ref{PD}c and~\ref{PD}d. Changes of the peak intensity for $x =$ 0.039, 0.043, 0.044, and 0.047 were measured at the (4,~0,~0) nuclear peak across the structural transition, which is associated with an extinction release\cite{Lester09, Kreyssig10,Lu_science_2014} (Fig.~\ref{PD}c). Measurements of extinction release as a structural order parameter can be very sensitive to the quality of the samples and can result in various shapes of order parameters as shown in Figures in Refs.~\onlinecite{Lester09, Kreyssig10,Lu_science_2014}. Therefore, even though extinction release is a result of a structural transition, measurements of extinction release should be marginally considered to represent structural order parameters. The AFM spin-density wave transitions for $x =$ 0.039, 0.043, 0.044, and 0.047 were measured at the (1,~0,~3) magnetic Bragg peak (Fig.~\ref{PD}d). Unlike the sharp transitions observed in $x =$ 0.028 (not shown), both order parameters show broad transitions, possibly due to a spread in compositions. 
In order to determine $T_\mathrm{S}$ and $T_\mathrm{N}$, we employ a power law fit with an additional Gaussian distribution of transition temperatures~\cite{Pajerowski13},

\begin{equation*}
I = A \int dt_\mathrm{N} \left[ \frac{1}{\sigma \sqrt{2\pi}}e^{-\frac{1}{2}\left( \frac{t_\mathrm{N}-T_\mathrm{N}}{\sigma} \right)^2} \left(\frac{t_\mathrm{N}-T}{t_\mathrm{N}}\right)^{2\beta} \right]
\label{eq1}
\end{equation*}
where $\sigma (=\sigma_\mathrm{S}, \sigma_\mathrm{N})$ is the standard deviation in temperature, $t_\mathrm{N}$. We present fit values of $T_\mathrm{S} = T_{\mathrm{S, fit}} \pm \sigma_\mathrm{S}$ and $T_\mathrm{N} = T_{\mathrm{N, fit}} \pm \sigma_\mathrm{N}$ in Table.~\ref{fit}. Considering the relative compositional uncertainty in each compound, the obtained temperature deviations in $T_\mathrm{S}$ and $T_\mathrm{N}$ are reasonable. For $x=$ 0.047 $\pm$ 0.002, we did not observe any changes in both structural and magnetic measurements and thus, we conclude that the magnetic and structural transitions are completely suppressed. We note that similar broad transitions in the AFM order parameters have been reported in superconducting Ba(Fe$_{1-x}$Ni$_{x}$)$_2$As$_2$ compounds and the authors conclude that it is associated with the spin glass state.\cite{Lu14} However, a spin glass state in Ba(Fe$_{0.957}$Cu$_{0.043}$)$_2$As$_2$ is less likely because of the relatively sharp magnetic ordering peak (Figs.~\ref{PD}a and b).

\begin{table}[b]
\caption {Fit values of the structural transition temperature ($T_\mathrm{S}$) and the AFM transition temperature ($T_\mathrm{N}$) for Ba(Fe$_{1-x}$Cu$_{x}$)$_2$As$_2$.}
\begin{ruledtabular}
\begin{tabular}{ |c|c|c|c|c| } 
Cu content & ~$T_\mathrm{S}$ (K)~ & ~$\sigma_\mathrm{S}$ (K)~ & ~$T_\mathrm{N}$ (K)~ & ~$\sigma_\mathrm{N}$ (K)~ \\
\hline
~~0.039 $\pm$ 0.002~~ & 42.7 & 4.0 & 31.3 & 3.3 \\
~~0.043 $\pm$ 0.002~~ & 33.3 & 3.7 & 26.2 & 4.5 \\ 
~~0.044 $\pm$ 0.001~~ & 16.9 & 6.6 & 12.3 & 6.7 \\ 
\end{tabular}
\label{fit}
\end{ruledtabular}
\end{table}

We measured the integrated intensity of the AFM Bragg peak for each sample and used the method described in Ref.~\onlinecite{Fernandes10} to estimate the ordered moment per Fe/Cu site, extrapolated to $T = 0$ K in each compound. The results are shown in the inset of Fig.~\ref{PD}e. The ordered moment decreases monotonically from $\sim$ 0.87 $\mu_\mathrm{B}$ at $x=0$~\cite{Huang08} to $\sim$ 0.15(4) $\mu_\mathrm{B}$ at $x=0.043$. The smooth reduction of the ordered moments with Cu substitution is similar to that with Co or Ni substitutions~\cite{Dai15,Fernandes10,Luo13}.

Altogether, we construct the phase diagram of Ba(Fe$_{1-x}$Cu$_{x}$)$_{2}$As$_{2}$ in Fig.~\ref{PD}e. Our phase diagram is consistent with the previous phase diagram\cite{canfield09, ni10}. Since the previous phase diagram was completed using bulk measurements, uncertainties in transition temperatures were large, especially in samples with higher Cu substitution level. By employing neutron diffraction, we significantly improve the uncertainties in $T_\mathrm{S}$ and $T_\mathrm{N}$. From Fig.~\ref{PD}e, it is readily seen that a putative antiferromagnetic quantum critical point exists at $0.044 < x \leq 0.047$.  

\subsection{Spin fluctuations at $\bm{E} \leq $ 14 meV: Comparison with $x =$ 0.028.}

\begin{figure}[t!]
\centering
\includegraphics[width=0.93\linewidth]{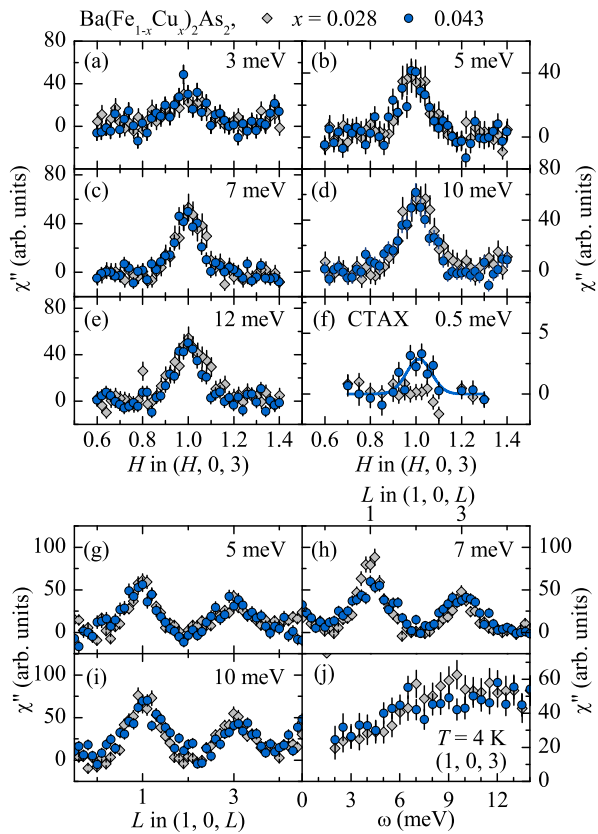} 
\caption{\rm{Spin fluctuation spectra at low energy transfers in Ba(Fe$_{1-x}$Cu$_x$)$_2$As$_2$ with $x=$ 0.028 and 0.043 at $T = 5$ K. Constant-energy $\bm{Q}$ scans along the orthorhombic $\bm{a}$ directions through (1,~0,~3) with energy transfers of (a) 3 meV, (b) 5 meV, (c) 7 meV, (d) 10 meV, (e) 12 meV, and (f) 0.5 meV. Constant-energy $\bm{Q}$ scans along the $\bm{c}$ directions with energy transfers of (g) 5 meV, (h) 7 meV, and (i) 10 meV. (j) Constant-momentum $E$ scans at (1,~0,~3). All measurements were done at HB3, except the data at $E = 0.5$ meV which was performed at CTAX. All data are corrected for the background and the Bose factor. The error bars indicate the statistical errors of one standard deviation.}}
\label{lowenergy}
\end{figure}

\begin{figure*}[t!]
\includegraphics[width=0.88\linewidth]{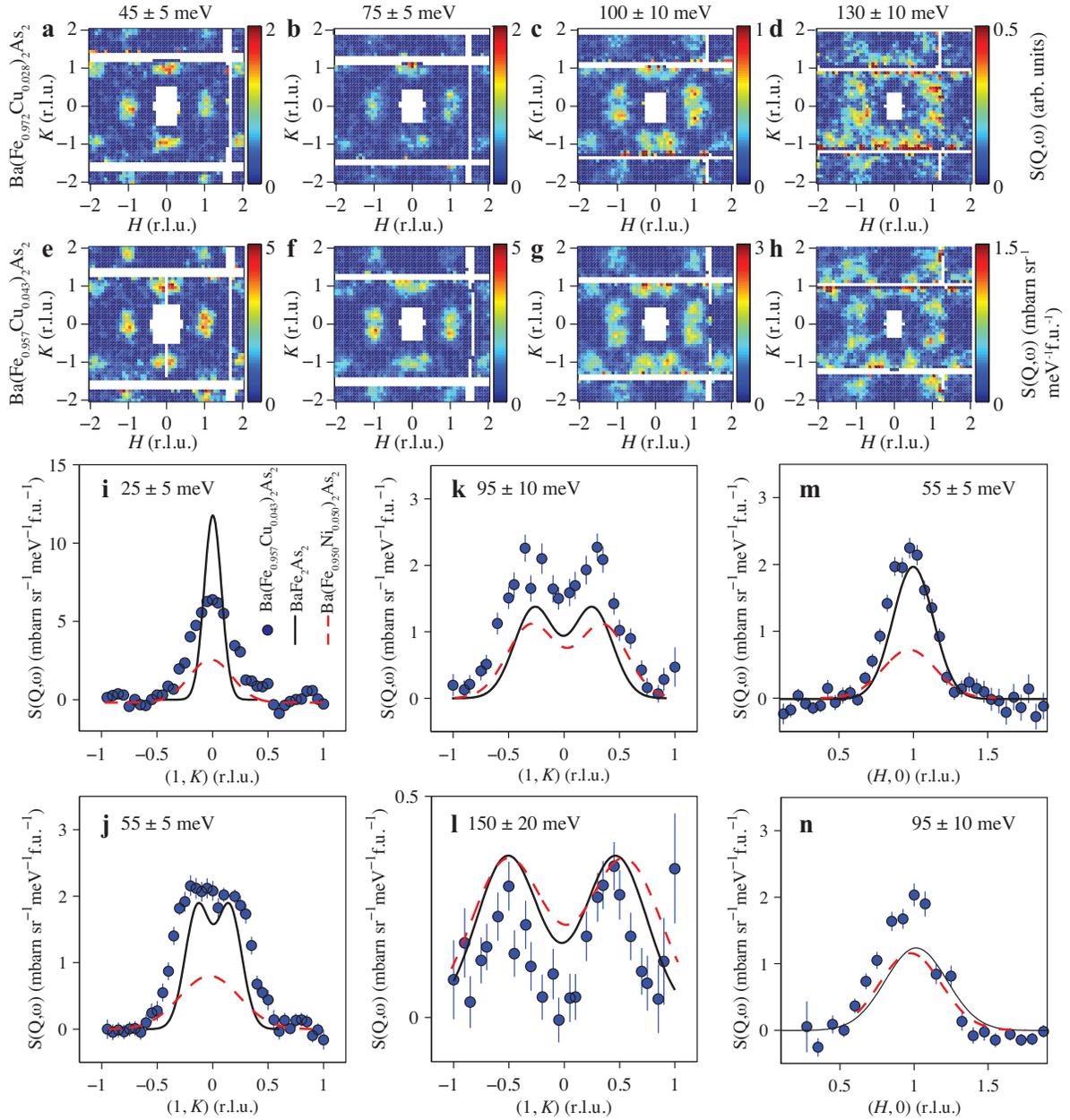} 
\caption{Two-dimensional images of spin fluctuations at (a) $45\pm5$ meV, (b) $75\pm5$ meV, (c) $100\pm10$ meV, (d) $130\pm10$ meV for Ba(Fe$_{0.972}$Cu$_{0.028}$)$_2$As$_2$ and (e) $45\pm5$ meV, (f) $75\pm5$ meV, (g) $100\pm10$ meV, (h) $130\pm10$ meV for Ba(Fe$_{0.957}$Cu$_{0.043}$)$_2$As$_2$. The images were measured at $T = 5$ K and obtained after the background subtraction. The color bars in (a)-(d) represent intensity in arbitrary units while the color bars in (e)-(h) indicate intensity in absolute units of mbarn sr$^{-1}$ meV$^{-1}$ f.u.$^{-1}$. (i)-(l) Constant-energy cut for Cu043 along the [1,~$K$] direction at $E=25\pm5$ meV, $55\pm5$ meV, $95\pm10$ meV, and $150\pm20$ meV, respectively. (m, n) Constant-energy cut for Cu043 along the [$H$,~0] direction at $E=55\pm5$ and $95\pm10$ meV, respectively. The solid lines and dashed lines in (i)-(n) represent the same cuts for the spin fluctuations of the parent BaFe$_2$As$_2$ ($T = 7$ K)~\cite{Harriger12} and superconducting Ba(Fe$_{0.95}$Ni$_{0.05}$)$_2$As$_2$ ($T = 5$ K)~\cite{Liu12, Luo13} in absolute units, respectively. The error bars indicate the statistical errors of one standard deviation.}
\label{highenergy}
\end{figure*}

In order to determine the effect of Cu substitution on the spin fluctuations, we present the results of inelastic neutron scattering in the low energy transfer regime measured at CTAX and HB3 in Fig.~\ref{lowenergy}. Figures~\ref{lowenergy}a-\ref{lowenergy}i show constant-$E$ $\bm{Q}$ scans along the orthorhombic $\bm{a}$ and $\bm{c}$ directions and Fig.~\ref{lowenergy}j shows constant-$\bm{Q}$ $E$ scans for Ba(Fe${_{0.972}}$Cu${_{0.028}}$)${_2}$As${_2}$ (`Cu028', $T_\mathrm{S} \approx 73$ K, $T_\mathrm{N} \approx 64$ K, no SC) and Ba(Fe${_{0.972}}$Cu${_{0.043}}$)${_2}$As${_2}$ (`Cu043', $T_\mathrm{S} \approx 33$ K, $T_\mathrm{N} \approx 26$ K, no SC). Since the measurements were performed on the same instruments with identical setups and the data are normalized by the total mass, we can directly compare the data for Cu028 and Cu043. We find that the spin gap is closed below $E = 0.5$ meV in Cu043 (Fig.~\ref{lowenergy}f) and the spin fluctuations at $E \leq 14$ meV are very similar in Cu028 and Cu043 as shown in Figures~\ref{lowenergy}a-\ref{lowenergy}e and~\ref{lowenergy}g-\ref{lowenergy}j. 
In the superconducting compounds, the spin correlations become weaker along the $\bm{c}$ direction, which is perpendicular to the FeAs plane, and the system becomes more quasi-two dimensional as it becomes more superconducting.\cite{Lumsden_2009, Harriger09, Tucker14} However, in our case with Cu substitution, the spin fluctuations along the $\bm{c}$ direction in both Cu028 and Cu043 remain unchanged, which indicates that the three-dimensional character of the spin fluctuations is preserved. It also contrasts with the width broadening observed in the static AFM peak along the $\bm{c}$ direction as seen in the previous section.

\subsection{Spin fluctuations at $\bm{E} \geq $ 14 meV: Comparison with $x =$ 0.028, the parent, and superconducting compounds.}

Now, we compare the spin fluctuations in  Cu043 and Cu028 at $E \geq 14$ meV. We present the results of the time-of-flight inelastic neutron scattering measurements obtained at ARCS in Fig.~\ref{highenergy}. We show two-dimensional (2D) images of the spin fluctuations in the ($H$,~$K$) plane at $E=45\pm5$, $75\pm5$, $100\pm10$, and $130\pm10$ meV for Cu028 in Figs.~\ref{highenergy}a-\ref{highenergy}d. The same set of 2D images for Cu043 is presented in Figs.~\ref{highenergy}e-\ref{highenergy}h. The 2D images look very similar between Cu028 and Cu043 and the analysis of the widths along $\bm{a}$ and $\bm{b}$ shows that the in-plane anisotropies for both Cu028 and Cu043 are comparable, which is consistent with the low energy results. The absolute unit conversion is not available for Cu028 while it is available for Cu043 (which is discussed later in this section) and thus the intensities of the spin fluctuations are not directly comparable between Cu028 and Cu043.

\begin{figure}[t!]
\includegraphics[width=0.95\linewidth]{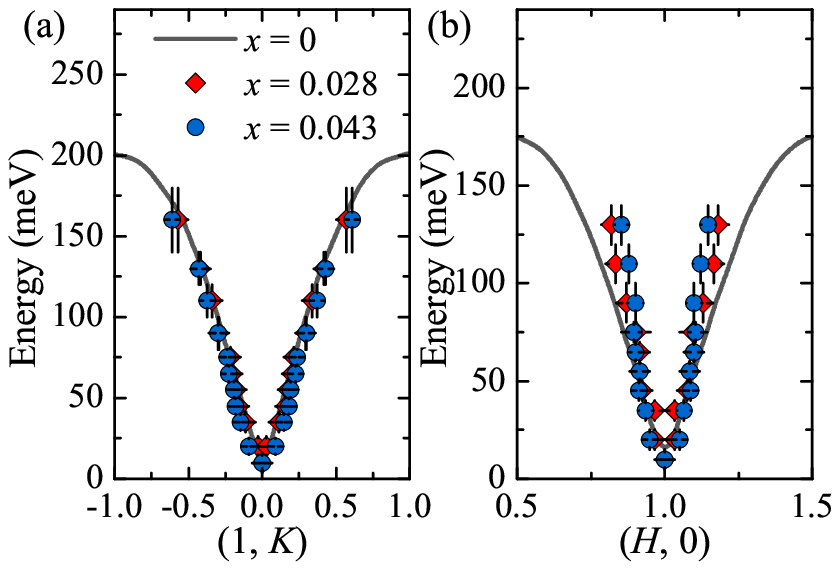}
\caption{{Dispersion of the spin fluctuations.} \rm {(a)} and {(b)} Dispersion of the spin fluctuations for the parent BaFe$_2$As$_2$ (lines, $T = 7$ K) from Ref.~\onlinecite{Harriger11}, Ba(Fe$_{0.972}$Cu$_{0.028}$)$_{2}$As$_{2}$ (diamonds), and Ba(Fe$_{0.957}$Cu$_{0.043}$)$_{2}$As$_{2}$ (circles) at $T = 5$ K.}
\label{dispersion}
\end{figure}

For further analysis, we cut through the 2D images similar to Figs.~\ref{highenergy}a-~\ref{highenergy}h along the [1,~$K$] and [$H$,~0] directions. Representative cuts are shown along the [1,~$K$] direction for $E=25\pm5$, $55\pm5$, $95\pm10$, and $150\pm20$ meV in Figs.~\ref{highenergy}i-~\ref{highenergy}l  and along the [$H$,~0] direction for $E=55\pm5$ and $95\pm10$ meV in Figs.~\ref{highenergy}m and~\ref{highenergy}n. We fit the cut data with a Gaussian function and determine the dispersion of the spin fluctuations along the two high symmetry directions and compare the dispersions between Cu028, Cu043, and the parent compound~\cite{Harriger11} in Figs.~\ref{dispersion}a and~\ref{dispersion}b. We find that at lower energy transfers ($E \leq 50$ meV), the dispersions for Cu028 and Cu043 look similar to the dispersion for the parent compound whereas at the higher energy transfers ($E > 50$ meV), the dispersion curve along the [$H$,~0] direction is stiffened.  In addition, the dispersion for Cu028 is very similar to that for Cu043 at all energy transfers, indicating that a small amount of Cu ($x= 0.028$) stiffens the dispersion along the orthorhombic $\bm{a}$ direction and the stiffening remains unchanged with more Cu substitution.

In order to quantify the changes in the spin fluctuations for Ba(Fe$_{0.957}$Cu$_{0.043}$)$_2$As$_2$, we fit the data using the Heisenberg $J_\mathrm{1a}-J_\mathrm{1b}-J_\mathrm{2}$ model\cite{Ewing08}. The neutron cross-section can be written as:

\begin{equation*}
\begin{split}
\frac{d^2 \sigma}{d\Omega dE} = &\frac{k_\mathrm{f}}{k_\mathrm{i}} \left( \frac{\gamma r_\mathrm{0}}{2} \right)^2
g^2f^2(Q)\exp(-2W) \\
&\times \sum_{\alpha\beta}\left(\delta_{\alpha\beta}-\Hat{Q}_{\alpha}\Hat{Q}_{\beta} \right)S^{\alpha\beta}(\bm{Q},E),
\label{eq2}
\end{split}
\end{equation*}

\noindent where $\frac{\gamma r_\mathrm{0}}{2}=72.65$ mbarn sr$^{-1}$, $g$ is the $g$-factor, $f(Q)$ is the form factor of iron, $\exp(-2W)$ is the Debye-Waller factor, $\Hat{Q}_{\alpha}$ is the $\alpha$ component of a unit vector, and the response function $S^{\alpha\beta}(\bm{Q},E)$ describes $\alpha\beta$ spin-spin correlations. Under the assumption that the transverse correlations only contribute to the spin wave cross-section, and finite excitation lifetimes can be described by a damped simple harmonic oscillator with the inverse lifetime $\Gamma$, 

\begin{equation*}
\begin{split}
& S^{yy}(\bm{Q},E)=S^{zz}(\bm{Q},E)= \\
& S_\mathrm{eff}\frac{(A_q-B_q)}{E_\mathrm{0}\left(1-\exp\left(\frac{-E}{k_\mathrm{B}T}\right)\right)}\frac{4}{\pi}
\frac{\Gamma E E_\mathrm{0}}{(E^2-E_\mathrm{0}^2)^2+4(\Gamma E)^2},
\label{eq3}
\end{split}
\end{equation*}

\noindent where $k_\mathrm{B}$ is the Boltzmann constant, $E_\mathrm{0}$ is the spin-wave energy, and $S_\mathrm{eff}$ is the effective spin. $A_q$ and $B_q$ are defined in the dispersion relations given by Refs.~\onlinecite{Zhao08,Ewing08,McQueeney08,Matan09,Diallo09}: $E(q)=\sqrt{A_q^2-B_q^2}$ with $A_q=2S[J_\mathrm{1a}(\cos(\pi K)-1)+J_\mathrm{1a}+J_\mathrm{c}+2J_2+J_\mathrm{s}]$ and $B_q=2S[J_\mathrm{1a}\cos(\pi H)+2J_2\cos(\pi H)\cos(\pi K)+J_\mathrm{c}\cos(\pi L)]$ with the single ion anisotropy constant $J_\mathrm{s}$. We employed the ResLib program\cite{Reslib} and the Tobyfit program\cite{Tobyfit} for our fit. We find that fitting using only low energy spectra ($E \leq 50$meV) yields  $SJ_\mathrm{1a} = 54.9 \pm 1.4$, $SJ_\mathrm{1b} = -5.9 \pm 2.7$, $SJ_\mathrm{2} = 17.3 \pm 1.2$, and $SJ_\mathrm{c} = 2.1 \pm 0.2$ meV. These values are similar to those measured in the FeAs family compounds.\cite{Zhao09, Harriger11, Ewings11} Fits for the entire energy range including $E > 50$ meV yield $SJ_\mathrm{1a} = 73.9 \pm 9.7$, $SJ_\mathrm{1b} = 10.4 \pm 2.5$, $SJ_2 = 18.2 \pm 2.2$, and $SJ_\mathrm{c} = 0.7 \pm 1.4$ meV. While only small changes are observed in $SJ_2$ and $SJ_\mathrm{c}$, a drastic modification occurs in the balance between $J_\mathrm{1a}$ and $J_\mathrm{1b}$. This may reflect the limitation of a simple spin wave model.

\begin{figure}[t!]
\includegraphics[width=0.95\linewidth]{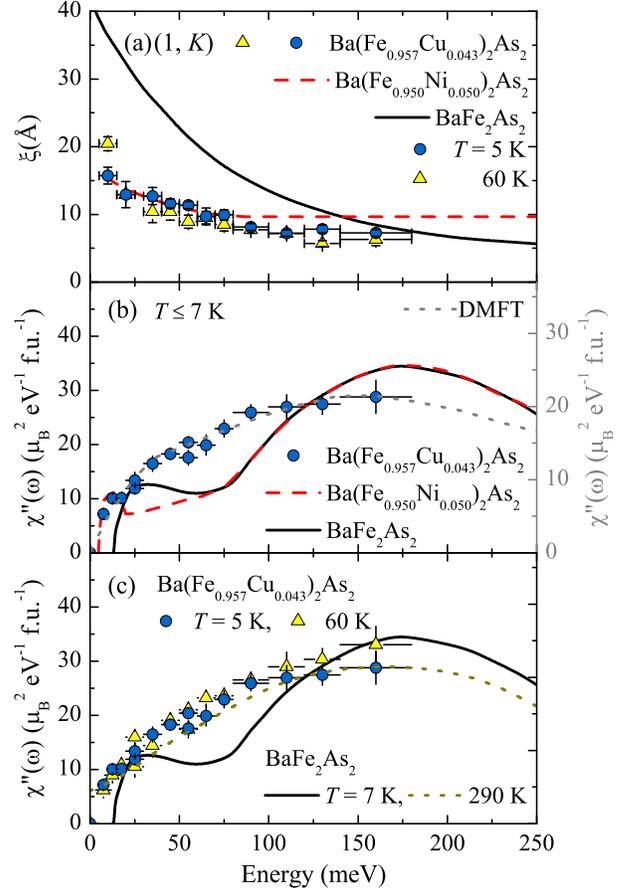} 
\caption{ (a) The dynamic spin-spin correlation length of Ba(Fe$_{0.957}$Cu$_{0.043}$)$_{2}$As$_{2}$ at $T = 5$ K (circles) and 60 K (triangles). (b) Energy dependence of the local susceptibility $\chi''(\omega)$ at $T = 5$ K (circles) and 60 K (triangles). The solid lines in (a) and (b) are for the parent BaFe$_2$As$_2$ compound at $T = 7$ K~\cite{Harriger12}. The dashed lines in (a) and (b) are for Ba(Fe$_{0.950}$Ni$_{0.050}$)$_2$As$_2$ at $T = 5$ K~\cite{Liu12, Luo13}. The dotted line in (b) is the DMFT calculation of $\chi''$ by Ref.~\onlinecite{Liu12}. (c) Temperature dependence of the local susceptibilities for Ba(Fe$_{0.957}$Cu$_{0.043}$)$_{2}$As$_{2}$ at $T = 5$ K (circles) and 60 K (triangles). The solid line and dotted lines in (c) are the local susceptibility for the parent BaFe$_2$As$_2$ compound\cite{Harriger12} at $T = 7$ K and 290 K.}
\label{sus}
\end{figure}

We normalize the time-of-flight data using a vanadium standard and plot them in absolute units of mbarn sr$^{-1}$ meV$^{-1}$ f.u.$^{-1}$ in Figs.~\ref{highenergy}e-\ref{highenergy}n and compare the data for Cu043 directly with those of the parent BaFe$_2$As$_2$ ($T_\mathrm{S} \approx T_\mathrm{N} \approx 137$ K, no SC)~\cite{Harriger11,Harriger12} and superconducting Ba(Fe$_{0.950}$Ni$_{0.050}$)$_2$As$_2$ (`Ni050', $T_\mathrm{S} \approx T_\mathrm{N} \approx 30$ K, $T_\mathrm{c}=20$ K)~\cite{Liu12,Luo13,Lu13}, which are plotted in absolute units in Figs.~\ref{highenergy}i-\ref{highenergy}n. We find that the spin fluctuations for Cu043 are significantly broader in momentum space than the fluctuations in the parent compound below $\approx 120$ meV but comparable to those in superconducting Ni050 for all energy transfers. We fit the cuts convolving the instrumental resolution and present the dynamic spin-spin correlation lengths for the parent (the solid line), Ni050 (the dashed line), and Cu043 (circles for $T = 4$K and diamonds for $T = 60$ K) in Fig.~\ref{sus}a. We find that the dynamic spin-spin correlation lengths are similar in Cu043 and Ni050 at all energies in contrast to the parent compound where the dynamic correlation length decreases rapidly with increasing energy. By comparing the data at $T = 5$ K (circles) and 60 K (diamonds) in Fig.~\ref{sus}a, we also find that the dynamic correlation length does not alter when Cu043 undergoes the AFM transition at $T_\mathrm{N} \approx 26$ K, suggesting no influence of the AFM ordering on the spin fluctuations in Cu043.

From Figs.~\ref{highenergy}i-\ref{highenergy}n, we find that the intensity of the spin fluctuations for Cu043 is stronger than that in Ni050 at $E < 150$ meV. This is in contrast to the result that the intensities of the spin fluctuations are indistinguishable for energies above 95 meV between the parent and Ni050. 
For quantitative comparison, Fig.~\ref{sus}b shows the momentum-integrated local dynamic susceptibility in absolute units, defined as $\chi''(\omega) = \int \chi''(\bm{q},\omega)d\bm{q}/ \int d\bm{q}$~\cite{Lester10}, where $\chi''(\bm{q},\omega) = (1/3)tr(\chi''_{\alpha \beta}(\bm{q},\omega))$, for the parent (solid line, $T = 7$ K), Ni050 (dashed line, $T = 5$ K), and Cu043 (circles, $T = 5$ K). As reported in Refs.~\onlinecite{Liu12, Luo13}, the overall shape of $\chi''(\omega)$ in the antiferromagnetically ordered superconducting compounds (i.e. Ni050 at $T = 5$ K in Fig.~\ref{sus}b) is very similar to the AFM spin fluctuations in the parent compound at $T = 7$ K, except for the reduction of $\chi''(\omega)$ at $E \leq 70$ meV and the spin-resonance below $T_\mathrm{c}$ (Fig.~\ref{sus}b). However, the local susceptibility of Cu043 (circles in Fig.~\ref{sus}b) is very different from that in both the parent at $T = 7$ K (solid line) and the Ni050 at $T = 5$ K (dashed line). 
To emphasize the difference, Fig.~\ref{sus}c shows the difference between the local susceptibility for Cu043 and the parent compound. We show the data at $T \leq 7$ K, the lowest measured temperature, and $T \approx 2T_\mathrm{N}$, which are $T = 60K$ for Cu043 and $T = 290$ K for the parent compound from Ref.~\onlinecite{Harriger12}. We find that the shapes of the local susceptibilities for Cu043 are very similar to the high-temperature spin fluctuations observed in the parent compound above its magnetic transition temperature ($T \geq T_\mathrm{N} \approx 140$ K) and consistent with the paramagnetic spin fluctuations (the dotted line in Fig.~\ref{sus}b), captured in the previous DMFT calculations (note, two different scales for experiments and the calculation)~\cite{Liu12}. 
Unlike the local susceptibility for the parent BaFe$_2$As$_2$ compound which shows an evolution from paramagnetic to antiferromagnetic spin fluctuations through $T_\mathrm{N}$ (Fig.~\ref{sus}c, Ref.~\onlinecite{Harriger12}), we find that the local susceptibility of Cu043 does not change through $T_\mathrm{N}$, implying no effect of the AFM ordering on the spin fluctuations.

For further inspection, we estimate the total fluctuating moments, defined as $\left< m^{2} \right> = (3\hbar / \pi) \int \chi''(\omega)d\omega / (1-\exp(-\hbar\omega / kT))$~\cite{Lester10}. Since our data are limited to energies below 160 meV, we assume that the local susceptibility at $E > 160$ meV in Cu043 follows the calculated form of the paramagnetic fluctuations from the DMFT calculation. We find $\left< m^{2} \right> = 3.48 \pm 0.17~\mu_B^2$ ($T = 5$ K) and $\left< m^{2} \right> = 3.62 \pm 0.15~\mu_B^2$ ($T = 60$ K) per Fe/Cu for Cu043. This is consistent with $S = 1/2$ for the magnetic moment of the spin $\left< m^{2} \right> = (g\mu_B)^2S(S+1)$ (where $g=2$) as observed in the parent compound\cite{Zhao09, Liu12,Harriger12}. Despite the much reduced ordered moment by 4.3 \% Cu substitution, the fluctuating moment remains similar to that in the parent compound, $\left< m^{2} \right> \approx 3.6~\mu_B^2$\cite{Harriger12}. 

\subsection{Antiferromagnetic Quantum Critical Point}

\begin{figure}[tI]
\centering
\includegraphics[width=0.95\linewidth]{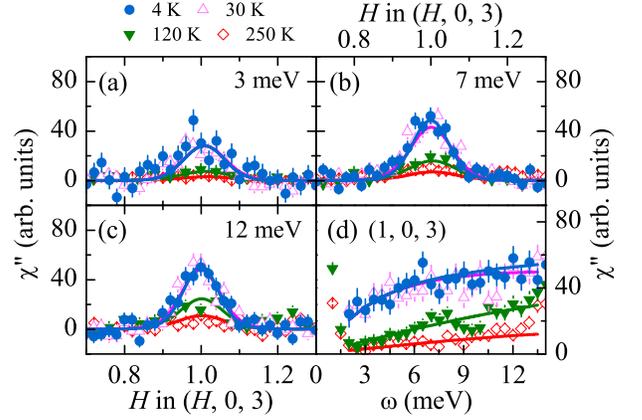}
\caption{ { The imaginary part of the dynamic susceptibility of Ba(Fe$_{0.957}$Cu$_{0.043}$)$_{2}$As$_{2}$ at selected temperature ($T =$ 4, 30, 120, and 250 K) and energies (3, 7, and 12 meV)} for Ba(Fe$_{0.957}$Cu$_{0.043}$)$_2$As$_2$. \rm {(a)-(d)} The data are obtained by subtracting the non-magnetic background and corrected for the Bose factor. Lines are the results of fits using the NAFL model as described in the text.}
\label{nafl}
\end{figure}

\begin{figure*}[ht!]
\includegraphics[width=0.8\linewidth]{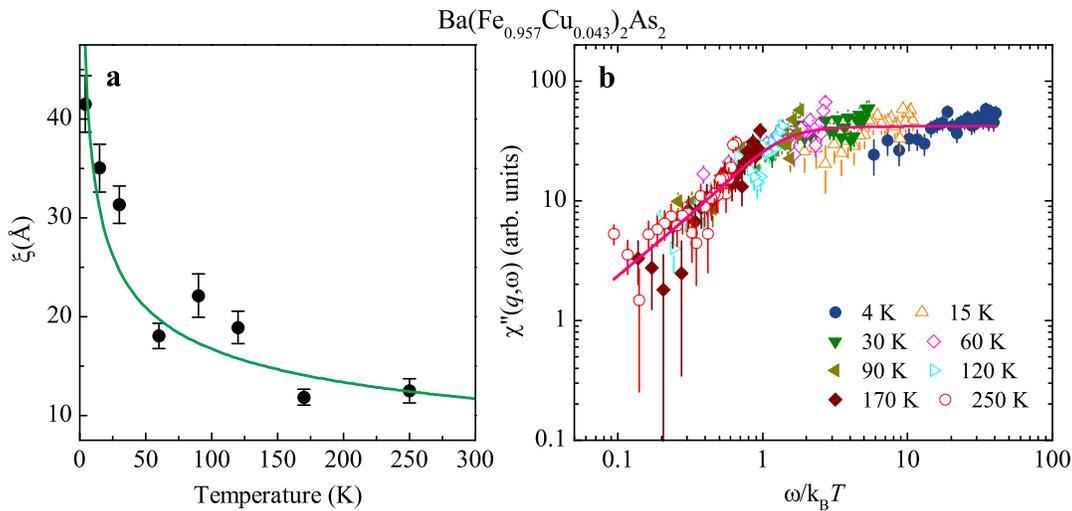} 
\caption{(a) The dynamic spin-spin correlation length obtained from the nearly antiferromagnetic Fermi liquid model as described in the text. The solid line demonstrates $\xi \propto T^{-0.32(5)}$. (b) The universal $\omega/T$ scaling  plot for the data with 2 meV $\leq E \leq$ 14 meV and 4 K $\leq T \leq$ 250 K. The solid line describes the best fit for $\chi''(\bm{q},\omega) \propto \arctan[a_1(\hbar\omega/k_BT) + a_2(\hbar\omega/k_BT)^2]$.}
\label{scaling}
\end{figure*}

To investigate the character of the spin fluctuations near the putative AFM QCP, low energy spin fluctuations were studied at several temperatures at the HB3 triple-axis spectrometer. We fit the paramagnon-like spin fluctuations in Cu043 using the nearly antiferromagnetic Fermi liquid model (NAFL) as describe by~\cite{Inosov09}:

\begin{equation*}
\chi''(\bm{q},\omega) = \frac{\chi_{0}\xi^{2}\Gamma \omega}{\omega^{2}+\Gamma^{2}\lbrack 1+\xi^{2}q^{2}\rbrack^{2}},
\label{eq1}
\end{equation*} 

\noindent where $\xi$ is the AFM correlation length and $\Gamma$ is the relaxation width due to the decay of spin waves into electron-hole pairs (Landau damping). The fits are shown with lines in Figs.~\ref{nafl}a-d for the data at selected temperature and energies; the NAFL model describes the low-energy spin fluctuations well in this compound.

We plot the fit values for the dynamic spin-spin correlation lengths $\xi$ in Fig.~\ref{scaling}a and find that the spin-spin correlation length increases rapidly as $T \rightarrow 0$. In the theory for magnetic quantum phase transitions for spin-density wave transitions, the correlation length scales as $\xi \propto T^{-\frac{d+z-2}{2z}}$ where $d$ is the dimension of the system and $z$ is the dynamical critical exponent that provides the scaling factors of space and time.~\cite{Hertz, Millis,Moriya,Sachdev_book} $z = 2$ for antiferromagnets and $d = 3$ for our system give $\xi \propto T^{-3/4}$. The power law fit (the solid line in Fig.~\ref{scaling}a) yields $\xi \propto T^{-0.32(5)}$, whose exponent is smaller than that in the theory. This is likely because Ba(Fe$_{0.957}$Cu$_{0.043}$)$_{2}$As$_{2}$ is not precisely at the QCP as we discuss more in the next section. We should note that more data points at different temperature are necessary for an accurate determination of the exponent.

We further test the scaling by directly plotting the imaginary part of the dynamical susceptibility, $\chi''(\bm{q},\omega)$. We find that the data between 4 K and 250 K collapse onto a single curve in Fig.~\ref{scaling}b; it confirms the $\omega/T$ scaling. In order to characterize the scaling, we attempt to fit our data with known functions, $f(\omega,k_BT)$, for the cuprate and the heavy-fermion superconductors.\cite{Aronson1995,Schroder1998,Keimer1992,Matsuda1993} We find that $\chi''(\bm{q},\omega) \propto \arctan[a_1(\hbar\omega/k_BT) + a_2(\hbar\omega/k_BT)^2]$ describes our data the best. This functional form was used for La$_{2-x}$Sr$_x$CuO$_{4+y}$\cite{Keimer1992,Matsuda1993} and is in agreement with the prediction of marginal Fermi liquid theory~\cite{Varma1989, Keimer1992}. The solid line in Fig.~\ref{scaling}b corresponds to the best fit for $\chi''(\bm{q},\omega) \propto \arctan[a_1(\hbar\omega/k_BT) + a_2(\hbar\omega/k_BT)^2]$ with $a_1 = 0.88(6)$ and $a_2 = 0.54(15)$.

\section{Discussion and Summary}

We have shown that the momentum-integrated local susceptibility for Ba(Fe$_{0.957}$Cu$_{0.043}$)$_2$As$_2$ ($T_\mathrm{S} = 33.3 \pm 3.7$ K, $T_\mathrm{N} = 26.2 \pm 4.5$ K, no SC) does not change when the system undergoes the antiferromagnetic phase transition and follows the form of the paramagnetic spin fluctuations. We argue that this is indicative of a weak (or absent) quasiparticle interaction with the spin fluctuations. Superconductivity and magnetism compete for the same quasiparticles in the iron pnictides\cite{Fernandes10}; in other words, quasiparticles contribute to the magnetism as well as to the Cooper pairs. In the parent BaFe$_2$As$_2$ compound, the quasiparticle interaction with the magnetism, specifically the AFM spin fluctuations, yields a clear evolution from paramagnetic (at $T \geq T_\mathrm{N}$) to antiferromagnetic fluctuations (at $T \leq T_\mathrm{N}$) (see Fig.~\ref{sus}c and Ref.~\onlinecite{Harriger12}). In superconducting Ba(Fe$_{1-x}$Ni$_x$)$_2$As$_2$ compounds, the AFM-like local susceptibility and the reduced fluctuating moments (see Fig. 16 in Ref.~\onlinecite{Luo13}) reflect the strong quasiparticle interaction through the AFM transition\cite{Liu12,Luo13}. In a similar vein, however, the paramagnon-like local susceptibility, the lack of temperature dependence of the local susceptibility, and preserved fluctuating moment for Ba(Fe$_{0.957}$Cu$_{0.043}$)$_2$As$_2$ provide evidence of a weak (or absent) quasiparticle interaction with the spin fluctuations. 
Two implications can be readily drawn from our observations. First, not only the ordered moments but also fluctuating moments are necessarily reduced via the strong quasiparticle interaction for high $T_\mathrm{c}$. While Cu substitution successfully decreases the ordered moment, the weak (or absent) quasiparticle interaction in Ba(Fe$_{0.957}$Cu$_{0.043}$)$_2$As$_2$ does not suppress the fluctuating moment; the preserved fluctuating moments are detrimental to the SC, which is consistent with the conventional effect of magnetism on superconductivity. Second, in terms of the fluctuation-mediated pairing mechanism, the quasiparticle interaction with the spin fluctuations represents the pairing strength. The introduction of the Cu substituent rapidly reduces the quasiparticle interaction and even when strong spin fluctuations exist at low temperature in this material, fluctuations can no longer act as the pairing medium. In this case, there may only be conventional electron-phonon interaction for the Cooper pairing, resulting in the possible existence of the bulk superconductivity at very low temperature ($T < 2$ K), which is consistent with the experiments and the prediction from the BCS theory for this compound\cite{ni10,Boeri08}. We should note that our results do not prove whether the spin fluctuations are the dominant mechanism for the Cooper pairing because different sorts of fluctuations, for instance orbital fluctuations, may be intricately connected to each other. 

Now we turn to a discussion on the quantum critical behavior in Ba(Fe$_{0.957}$Cu$_{0.043}$)$_2$As$_2$. We have shown the details of the spin fluctuations at several temperatures including a rapid increase of the dynamic spin-spin correlation length as $T \rightarrow 0$ together with $\omega/T$ scaling in Ba(Fe$_{0.957}$Cu$_{0.043}$)$_2$As$_2$ as supporting evidences of an antiferromagnetic quantum critical point (AFM QCP). The dynamic spin-spin correlation length at the QCP is expected to diverge as $\xi \propto T^{-3/4}$ in Ba(Fe$_{0.957}$Cu$_{0.043}$)$_2$As$_2$\cite{Hertz, Millis,Moriya,Sachdev_book}. However, our sample is not exactly at the QCP but proximal to it so that the dynamic correlation length should saturate at a finite value, resulting in a smaller effective exponent as we observe. While our results support the existence of the antiferromagnetic quantum critical point (AFM QCP), as mentioned earlier, the observed broad AFM order parameter in this compound suggests the possibility of a spin glass state in Ba(Fe$_{0.957}$Cu$_{0.043}$)$_2$As$_2$. Further, broadening of the (1,~0,~3) magnetic peak in Cu043 compared to that in Cu028 may also support the emergence of a spin glass state with increasing Cu substitution while the ordering is still long-ranged in the current sample. Therefore, we cannot exclude a possible spin glass quantum critical point. Interestingly, neither the Fermi-liquid description of the AFM QCP\cite{Hertz, Millis,Moriya,Sachdev_book} nor the theoretical models for the spin glass QCP\cite{sachdev95,Segupta95,Georges01} predict $\omega/T$ scaling. In addition, the possible spin glass state may be intrinsically associated with disorder that can, itself, play an important role in the observed $\omega/T$ scaling. In some of the heavy-fermion materials, such as UCu$_{5-x}$Pd$_x$ compounds, studies have attributed the non-Fermi liquid behavior, including $\omega/T$ scaling, to disorder, which is unrelated to the quantum criticality.\cite{Bernal1995, Miranda1996, laughlin04,Lohneysen_2007} In Cu substituted BaFe$_2$As$_2$ compounds, recent DFT and neutron diffraction studies\cite{Kim12} and our observation of broad AFM peaks and order parameters suggest that the disorder introduced by Cu substitution may play an important role in the quantum critical behavior in Ba(Fe$_{0.957}$Cu$_{0.043}$)$_2$As$_2$. We note that the disorder is also expected to impact superconductivity and may explain the absence of superconductivity in this compound.\cite{Kim12, Vavilov11, Fernandes_12}
Although further studies on both theory and experiment are necessary to understand the character and mechanism of the QCP, our experimental observation nevertheless demonstrates the existence of a quantum critical point in Ba(Fe$_{1-x}$Cu$_{x}$)$_2$As$_2$.


\begin{acknowledgments} 

The work at the Lawrence Berkeley National Laboratory was supported by the U.S. Department of Energy (DOE), Office of Basic Energy Sciences, Materials Sciences and Engineering Division, under Contract No. DE-AC02-05CH11231. The work at the Ames Laboratory was supported by the Department of Energy-Basic Energy Sciences under Contract No. DE-AC02-07CH11358. Research conducted at ORNL's HFIR and SNS was sponsored by the Scientific User Facilities Division, Office of Basic Energy Sciences,US Department of Energy.
\end{acknowledgments}

\bibliographystyle{apsrev}
\bibliography{cu_ins}

\begin{thebibliography}{62}
\expandafter\ifx\csname natexlab\endcsname\relax\def\natexlab#1{#1}\fi
\expandafter\ifx\csname bibnamefont\endcsname\relax
  \def\bibnamefont#1{#1}\fi
\expandafter\ifx\csname bibfnamefont\endcsname\relax
  \def\bibfnamefont#1{#1}\fi
\expandafter\ifx\csname citenamefont\endcsname\relax
  \def\citenamefont#1{#1}\fi
\expandafter\ifx\csname url\endcsname\relax
  \def\url#1{\texttt{#1}}\fi
\expandafter\ifx\csname urlprefix\endcsname\relax\def\urlprefix{URL }\fi
\providecommand{\bibinfo}[2]{#2}
\providecommand{\eprint}[2][]{\url{#2}}

\bibitem[{\citenamefont{Harriger et~al.}(2012)\citenamefont{Harriger, Liu, Luo,
  Ewings, Frost, Perring, and Dai}}]{Harriger12}
\bibinfo{author}{\bibfnamefont{L.~W.} \bibnamefont{Harriger}},
  \bibinfo{author}{\bibfnamefont{M.}~\bibnamefont{Liu}},
  \bibinfo{author}{\bibfnamefont{H.}~\bibnamefont{Luo}},
  \bibinfo{author}{\bibfnamefont{R.~A.} \bibnamefont{Ewings}},
  \bibinfo{author}{\bibfnamefont{C.~D.} \bibnamefont{Frost}},
  \bibinfo{author}{\bibfnamefont{T.~G.} \bibnamefont{Perring}},
  \bibnamefont{and} \bibinfo{author}{\bibfnamefont{P.}~\bibnamefont{Dai}},
  \bibinfo{journal}{Phys. Rev. B} \textbf{\bibinfo{volume}{86}},
  \bibinfo{pages}{140403} (\bibinfo{year}{2012}).

\bibitem[{\citenamefont{Liu et~al.}(2012)\citenamefont{Liu, Herriger, Luo,
  Wang, Ewings, Guidi, Park, Haule, Kotliar, Hayden et~al.}}]{Liu12}
\bibinfo{author}{\bibfnamefont{M.}~\bibnamefont{Liu}},
  \bibinfo{author}{\bibfnamefont{L.~W.} \bibnamefont{Herriger}},
  \bibinfo{author}{\bibfnamefont{H.}~\bibnamefont{Luo}},
  \bibinfo{author}{\bibfnamefont{M.}~\bibnamefont{Wang}},
  \bibinfo{author}{\bibfnamefont{R.~A.} \bibnamefont{Ewings}},
  \bibinfo{author}{\bibfnamefont{T.}~\bibnamefont{Guidi}},
  \bibinfo{author}{\bibfnamefont{H.}~\bibnamefont{Park}},
  \bibinfo{author}{\bibfnamefont{K.}~\bibnamefont{Haule}},
  \bibinfo{author}{\bibfnamefont{G.}~\bibnamefont{Kotliar}},
  \bibinfo{author}{\bibfnamefont{S.~M.} \bibnamefont{Hayden}},
  \bibnamefont{et~al.}, \bibinfo{journal}{Nature Phys.}
  \textbf{\bibinfo{volume}{8}}, \bibinfo{pages}{376} (\bibinfo{year}{2012}).

\bibitem[{\citenamefont{Luo et~al.}(2013)\citenamefont{Luo, Lu, Zhang, Wang,
  Goremychkin, Adroja, Danilkin, Deng, Yamani, and Dai}}]{Luo13}
\bibinfo{author}{\bibfnamefont{H.}~\bibnamefont{Luo}},
  \bibinfo{author}{\bibfnamefont{X.}~\bibnamefont{Lu}},
  \bibinfo{author}{\bibfnamefont{R.}~\bibnamefont{Zhang}},
  \bibinfo{author}{\bibfnamefont{M.}~\bibnamefont{Wang}},
  \bibinfo{author}{\bibfnamefont{E.~A.} \bibnamefont{Goremychkin}},
  \bibinfo{author}{\bibfnamefont{D.~T.} \bibnamefont{Adroja}},
  \bibinfo{author}{\bibfnamefont{S.}~\bibnamefont{Danilkin}},
  \bibinfo{author}{\bibfnamefont{G.}~\bibnamefont{Deng}},
  \bibinfo{author}{\bibfnamefont{Z.}~\bibnamefont{Yamani}}, \bibnamefont{and}
  \bibinfo{author}{\bibfnamefont{P.}~\bibnamefont{Dai}},
  \bibinfo{journal}{Phys. Rev. B} \textbf{\bibinfo{volume}{88}},
  \bibinfo{pages}{144516} (\bibinfo{year}{2013}).

\bibitem[{\citenamefont{Monthoux et~al.}(2007)\citenamefont{Monthoux, Pines,
  and Lonzarich}}]{Monthoux07}
\bibinfo{author}{\bibfnamefont{P.}~\bibnamefont{Monthoux}},
  \bibinfo{author}{\bibfnamefont{D.}~\bibnamefont{Pines}}, \bibnamefont{and}
  \bibinfo{author}{\bibfnamefont{G.~G.} \bibnamefont{Lonzarich}},
  \bibinfo{journal}{Nature} \textbf{\bibinfo{volume}{450}},
  \bibinfo{pages}{1177} (\bibinfo{year}{2007}).

\bibitem[{\citenamefont{Wang and Lee}(2011)}]{Wang11}
\bibinfo{author}{\bibfnamefont{F.}~\bibnamefont{Wang}} \bibnamefont{and}
  \bibinfo{author}{\bibfnamefont{D.-H.} \bibnamefont{Lee}},
  \bibinfo{journal}{Science} \textbf{\bibinfo{volume}{332}},
  \bibinfo{pages}{200} (\bibinfo{year}{2011}).

\bibitem[{\citenamefont{Scalapino}(2012)}]{Scalapino12}
\bibinfo{author}{\bibfnamefont{D.~J.} \bibnamefont{Scalapino}},
  \bibinfo{journal}{Rev. Mod. Phys.} \textbf{\bibinfo{volume}{84}},
  \bibinfo{pages}{1383} (\bibinfo{year}{2012}).

\bibitem[{\citenamefont{Hosono and Kuroki}(2015)}]{Hosono15}
\bibinfo{author}{\bibfnamefont{H.}~\bibnamefont{Hosono}} \bibnamefont{and}
  \bibinfo{author}{\bibfnamefont{K.}~\bibnamefont{Kuroki}},
  \bibinfo{journal}{Physica C} \textbf{\bibinfo{volume}{514}},
  \bibinfo{pages}{399} (\bibinfo{year}{2015}).

\bibitem[{\citenamefont{Johnston}(2010)}]{Johnston10}
\bibinfo{author}{\bibfnamefont{D.}~\bibnamefont{Johnston}},
  \bibinfo{journal}{Adv. Phys.} \textbf{\bibinfo{volume}{59}},
  \bibinfo{pages}{803} (\bibinfo{year}{2010}).

\bibitem[{\citenamefont{Canfield and Bud'ko}(2010)}]{canfield_feas_2010}
\bibinfo{author}{\bibfnamefont{P.~C.} \bibnamefont{Canfield}} \bibnamefont{and}
  \bibinfo{author}{\bibfnamefont{S.~L.} \bibnamefont{Bud'ko}},
  \bibinfo{journal}{Annu. Rev. Condens. Matter Phys.}
  \textbf{\bibinfo{volume}{1}}, \bibinfo{pages}{27} (\bibinfo{year}{2010}).

\bibitem[{\citenamefont{Dai}(2015)}]{Dai15}
\bibinfo{author}{\bibfnamefont{P.}~\bibnamefont{Dai}}, \bibinfo{journal}{Rev.
  Mod. Phys.} \textbf{\bibinfo{volume}{87}}, \bibinfo{pages}{855}
  (\bibinfo{year}{2015}).

\bibitem[{\citenamefont{Canfield et~al.}(2009)\citenamefont{Canfield, Bud'ko,
  Ni, Yan, and Kracher}}]{canfield09}
\bibinfo{author}{\bibfnamefont{P.~C.} \bibnamefont{Canfield}},
  \bibinfo{author}{\bibfnamefont{S.~L.} \bibnamefont{Bud'ko}},
  \bibinfo{author}{\bibfnamefont{N.}~\bibnamefont{Ni}},
  \bibinfo{author}{\bibfnamefont{J.~Q.} \bibnamefont{Yan}}, \bibnamefont{and}
  \bibinfo{author}{\bibfnamefont{A.}~\bibnamefont{Kracher}},
  \bibinfo{journal}{Phys. Rev. B} \textbf{\bibinfo{volume}{80}},
  \bibinfo{pages}{060501} (\bibinfo{year}{2009}).

\bibitem[{\citenamefont{Ni et~al.}(2008)\citenamefont{Ni, Tillman, Yan,
  Kracher, Hannahs, Bud'ko, and Canfield}}]{ni08}
\bibinfo{author}{\bibfnamefont{N.}~\bibnamefont{Ni}},
  \bibinfo{author}{\bibfnamefont{M.~E.} \bibnamefont{Tillman}},
  \bibinfo{author}{\bibfnamefont{J.-Q.} \bibnamefont{Yan}},
  \bibinfo{author}{\bibfnamefont{A.}~\bibnamefont{Kracher}},
  \bibinfo{author}{\bibfnamefont{S.~T.} \bibnamefont{Hannahs}},
  \bibinfo{author}{\bibfnamefont{S.~L.} \bibnamefont{Bud'ko}},
  \bibnamefont{and} \bibinfo{author}{\bibfnamefont{P.~C.}
  \bibnamefont{Canfield}}, \bibinfo{journal}{Phys. Rev. B}
  \textbf{\bibinfo{volume}{78}}, \bibinfo{pages}{214515}
  (\bibinfo{year}{2008}).

\bibitem[{\citenamefont{Li et~al.}(2009)\citenamefont{Li, Luo, Wang, Chen, Ren,
  Tao, Li, Lin, He, Zhu et~al.}}]{li_superconductivity_2009}
\bibinfo{author}{\bibfnamefont{L.~J.} \bibnamefont{Li}},
  \bibinfo{author}{\bibfnamefont{Y.~K.} \bibnamefont{Luo}},
  \bibinfo{author}{\bibfnamefont{Q.~B.} \bibnamefont{Wang}},
  \bibinfo{author}{\bibfnamefont{H.}~\bibnamefont{Chen}},
  \bibinfo{author}{\bibfnamefont{Z.}~\bibnamefont{Ren}},
  \bibinfo{author}{\bibfnamefont{Q.}~\bibnamefont{Tao}},
  \bibinfo{author}{\bibfnamefont{Y.~K.} \bibnamefont{Li}},
  \bibinfo{author}{\bibfnamefont{X.}~\bibnamefont{Lin}},
  \bibinfo{author}{\bibfnamefont{M.}~\bibnamefont{He}},
  \bibinfo{author}{\bibfnamefont{Z.~W.} \bibnamefont{Zhu}},
  \bibnamefont{et~al.}, \bibinfo{journal}{New J. Phys.}
  \textbf{\bibinfo{volume}{11}}, \bibinfo{pages}{025008}
  (\bibinfo{year}{2009}).

\bibitem[{\citenamefont{Ni et~al.}(2010)\citenamefont{Ni, Thaler, Yan, Kracher,
  Colombier, Bud'ko, Canfield, and Hannahs}}]{ni10}
\bibinfo{author}{\bibfnamefont{N.}~\bibnamefont{Ni}},
  \bibinfo{author}{\bibfnamefont{A.}~\bibnamefont{Thaler}},
  \bibinfo{author}{\bibfnamefont{J.~Q.} \bibnamefont{Yan}},
  \bibinfo{author}{\bibfnamefont{A.}~\bibnamefont{Kracher}},
  \bibinfo{author}{\bibfnamefont{E.}~\bibnamefont{Colombier}},
  \bibinfo{author}{\bibfnamefont{S.~L.} \bibnamefont{Bud'ko}},
  \bibinfo{author}{\bibfnamefont{P.~C.} \bibnamefont{Canfield}},
  \bibnamefont{and} \bibinfo{author}{\bibfnamefont{S.~T.}
  \bibnamefont{Hannahs}}, \bibinfo{journal}{Phys. Rev. B}
  \textbf{\bibinfo{volume}{82}}, \bibinfo{pages}{024519}
  (\bibinfo{year}{2010}).

\bibitem[{\citenamefont{Fernandes et~al.}(2010)\citenamefont{Fernandes, Pratt,
  Tian, Zarestky, Kreyssig, Nandi, Kim, Thaler, Ni, Canfield
  et~al.}}]{Fernandes10}
\bibinfo{author}{\bibfnamefont{R.~M.} \bibnamefont{Fernandes}},
  \bibinfo{author}{\bibfnamefont{D.~K.} \bibnamefont{Pratt}},
  \bibinfo{author}{\bibfnamefont{W.}~\bibnamefont{Tian}},
  \bibinfo{author}{\bibfnamefont{J.}~\bibnamefont{Zarestky}},
  \bibinfo{author}{\bibfnamefont{A.}~\bibnamefont{Kreyssig}},
  \bibinfo{author}{\bibfnamefont{S.}~\bibnamefont{Nandi}},
  \bibinfo{author}{\bibfnamefont{M.~G.} \bibnamefont{Kim}},
  \bibinfo{author}{\bibfnamefont{A.}~\bibnamefont{Thaler}},
  \bibinfo{author}{\bibfnamefont{N.}~\bibnamefont{Ni}},
  \bibinfo{author}{\bibfnamefont{P.~C.} \bibnamefont{Canfield}},
  \bibnamefont{et~al.}, \bibinfo{journal}{Phys. Rev. B}
  \textbf{\bibinfo{volume}{81}}, \bibinfo{pages}{140501}
  (\bibinfo{year}{2010}).

\bibitem[{\citenamefont{Lu et~al.}(2013)\citenamefont{Lu, Gretarsson, Zhang,
  Liu, Luo, Tian, Laver, Yamani, Kim, Nevidomskyy et~al.}}]{Lu13}
\bibinfo{author}{\bibfnamefont{X.}~\bibnamefont{Lu}},
  \bibinfo{author}{\bibfnamefont{H.}~\bibnamefont{Gretarsson}},
  \bibinfo{author}{\bibfnamefont{R.}~\bibnamefont{Zhang}},
  \bibinfo{author}{\bibfnamefont{X.}~\bibnamefont{Liu}},
  \bibinfo{author}{\bibfnamefont{H.}~\bibnamefont{Luo}},
  \bibinfo{author}{\bibfnamefont{W.}~\bibnamefont{Tian}},
  \bibinfo{author}{\bibfnamefont{M.}~\bibnamefont{Laver}},
  \bibinfo{author}{\bibfnamefont{Z.}~\bibnamefont{Yamani}},
  \bibinfo{author}{\bibfnamefont{Y.-J.} \bibnamefont{Kim}},
  \bibinfo{author}{\bibfnamefont{A.~H.} \bibnamefont{Nevidomskyy}},
  \bibnamefont{et~al.}, \bibinfo{journal}{Phys. Rev. Lett.}
  \textbf{\bibinfo{volume}{110}}, \bibinfo{pages}{257001}
  (\bibinfo{year}{2013}).

\bibitem[{\citenamefont{Lu et~al.}(2014{\natexlab{a}})\citenamefont{Lu, Tam,
  Zhang, Luo, Wang, Zhang, Harriger, Keller, Keimer, Regnault et~al.}}]{Lu14}
\bibinfo{author}{\bibfnamefont{X.}~\bibnamefont{Lu}},
  \bibinfo{author}{\bibfnamefont{D.~W.} \bibnamefont{Tam}},
  \bibinfo{author}{\bibfnamefont{C.}~\bibnamefont{Zhang}},
  \bibinfo{author}{\bibfnamefont{H.}~\bibnamefont{Luo}},
  \bibinfo{author}{\bibfnamefont{M.}~\bibnamefont{Wang}},
  \bibinfo{author}{\bibfnamefont{R.}~\bibnamefont{Zhang}},
  \bibinfo{author}{\bibfnamefont{L.~W.} \bibnamefont{Harriger}},
  \bibinfo{author}{\bibfnamefont{T.}~\bibnamefont{Keller}},
  \bibinfo{author}{\bibfnamefont{B.}~\bibnamefont{Keimer}},
  \bibinfo{author}{\bibfnamefont{L.-P.} \bibnamefont{Regnault}},
  \bibnamefont{et~al.}, \bibinfo{journal}{Phys. Rev. B}
  \textbf{\bibinfo{volume}{90}}, \bibinfo{pages}{024509}
  (\bibinfo{year}{2014}{\natexlab{a}}).

\bibitem[{\citenamefont{Hu et~al.}(2015)\citenamefont{Hu, Lu, Zhang, Luo, Li,
  Wang, Chen, Han, Banjara, Sapkota et~al.}}]{Hu15}
\bibinfo{author}{\bibfnamefont{D.}~\bibnamefont{Hu}},
  \bibinfo{author}{\bibfnamefont{X.}~\bibnamefont{Lu}},
  \bibinfo{author}{\bibfnamefont{W.}~\bibnamefont{Zhang}},
  \bibinfo{author}{\bibfnamefont{H.}~\bibnamefont{Luo}},
  \bibinfo{author}{\bibfnamefont{S.}~\bibnamefont{Li}},
  \bibinfo{author}{\bibfnamefont{P.}~\bibnamefont{Wang}},
  \bibinfo{author}{\bibfnamefont{G.}~\bibnamefont{Chen}},
  \bibinfo{author}{\bibfnamefont{F.}~\bibnamefont{Han}},
  \bibinfo{author}{\bibfnamefont{S.~R.} \bibnamefont{Banjara}},
  \bibinfo{author}{\bibfnamefont{A.}~\bibnamefont{Sapkota}},
  \bibnamefont{et~al.}, \bibinfo{journal}{Phys. Rev. Lett.}
  \textbf{\bibinfo{volume}{114}}, \bibinfo{pages}{157002}
  (\bibinfo{year}{2015}).

\bibitem[{\citenamefont{Kasahara et~al.}(2010)\citenamefont{Kasahara,
  Shibauchi, Hashimoto, Ikada, Tonegawa, Okazaki, Shishido, Ikeda, Takeya,
  Hirata et~al.}}]{Kasahara10}
\bibinfo{author}{\bibfnamefont{S.}~\bibnamefont{Kasahara}},
  \bibinfo{author}{\bibfnamefont{T.}~\bibnamefont{Shibauchi}},
  \bibinfo{author}{\bibfnamefont{K.}~\bibnamefont{Hashimoto}},
  \bibinfo{author}{\bibfnamefont{K.}~\bibnamefont{Ikada}},
  \bibinfo{author}{\bibfnamefont{S.}~\bibnamefont{Tonegawa}},
  \bibinfo{author}{\bibfnamefont{R.}~\bibnamefont{Okazaki}},
  \bibinfo{author}{\bibfnamefont{H.}~\bibnamefont{Shishido}},
  \bibinfo{author}{\bibfnamefont{H.}~\bibnamefont{Ikeda}},
  \bibinfo{author}{\bibfnamefont{H.}~\bibnamefont{Takeya}},
  \bibinfo{author}{\bibfnamefont{K.}~\bibnamefont{Hirata}},
  \bibnamefont{et~al.}, \bibinfo{journal}{Phys. Rev. B}
  \textbf{\bibinfo{volume}{81}}, \bibinfo{pages}{184519}
  (\bibinfo{year}{2010}).

\bibitem[{\citenamefont{Nakai et~al.}(2010)\citenamefont{Nakai, Iye, Kitagawa,
  Ishida, Ikeda, Kasahara, Shishido, Shibauchi, Matsuda, and
  Terashima}}]{Nakai10}
\bibinfo{author}{\bibfnamefont{Y.}~\bibnamefont{Nakai}},
  \bibinfo{author}{\bibfnamefont{T.}~\bibnamefont{Iye}},
  \bibinfo{author}{\bibfnamefont{S.}~\bibnamefont{Kitagawa}},
  \bibinfo{author}{\bibfnamefont{K.}~\bibnamefont{Ishida}},
  \bibinfo{author}{\bibfnamefont{H.}~\bibnamefont{Ikeda}},
  \bibinfo{author}{\bibfnamefont{S.}~\bibnamefont{Kasahara}},
  \bibinfo{author}{\bibfnamefont{H.}~\bibnamefont{Shishido}},
  \bibinfo{author}{\bibfnamefont{T.}~\bibnamefont{Shibauchi}},
  \bibinfo{author}{\bibfnamefont{Y.}~\bibnamefont{Matsuda}}, \bibnamefont{and}
  \bibinfo{author}{\bibfnamefont{T.}~\bibnamefont{Terashima}},
  \bibinfo{journal}{Phys. Rev. Lett.} \textbf{\bibinfo{volume}{105}},
  \bibinfo{pages}{107003} (\bibinfo{year}{2010}).

\bibitem[{\citenamefont{Hashimoto et~al.}(2012)\citenamefont{Hashimoto, Cho,
  Shibauchi, Kasahara, Mizukami, Katsumata, Tsuruhara, Terashima, Ikeda,
  Tanatar et~al.}}]{Hashimoto_2012}
\bibinfo{author}{\bibfnamefont{K.}~\bibnamefont{Hashimoto}},
  \bibinfo{author}{\bibfnamefont{K.}~\bibnamefont{Cho}},
  \bibinfo{author}{\bibfnamefont{T.}~\bibnamefont{Shibauchi}},
  \bibinfo{author}{\bibfnamefont{S.}~\bibnamefont{Kasahara}},
  \bibinfo{author}{\bibfnamefont{Y.}~\bibnamefont{Mizukami}},
  \bibinfo{author}{\bibfnamefont{R.}~\bibnamefont{Katsumata}},
  \bibinfo{author}{\bibfnamefont{Y.}~\bibnamefont{Tsuruhara}},
  \bibinfo{author}{\bibfnamefont{T.}~\bibnamefont{Terashima}},
  \bibinfo{author}{\bibfnamefont{H.}~\bibnamefont{Ikeda}},
  \bibinfo{author}{\bibfnamefont{M.~A.} \bibnamefont{Tanatar}},
  \bibnamefont{et~al.}, \bibinfo{journal}{Science}
  \textbf{\bibinfo{volume}{22}}, \bibinfo{pages}{1554} (\bibinfo{year}{2012}).

\bibitem[{\citenamefont{L\"ohneysen et~al.}(2007)\citenamefont{L\"ohneysen,
  Rosch, Vojta, and W\"olfle}}]{Lohneysen_2007}
\bibinfo{author}{\bibfnamefont{H.~v.} \bibnamefont{L\"ohneysen}},
  \bibinfo{author}{\bibfnamefont{A.}~\bibnamefont{Rosch}},
  \bibinfo{author}{\bibfnamefont{M.}~\bibnamefont{Vojta}}, \bibnamefont{and}
  \bibinfo{author}{\bibfnamefont{P.}~\bibnamefont{W\"olfle}},
  \bibinfo{journal}{Rev. Mod. Phys.} \textbf{\bibinfo{volume}{79}},
  \bibinfo{pages}{1015} (\bibinfo{year}{2007}).

\bibitem[{\citenamefont{Sachdev}(2011)}]{Sachdev_book}
\bibinfo{author}{\bibfnamefont{S.}~\bibnamefont{Sachdev}},
  \emph{\bibinfo{title}{Quantum Phase Transitions}}
  (\bibinfo{publisher}{Cambridge University Press}, \bibinfo{address}{London},
  \bibinfo{year}{2011}).

\bibitem[{\citenamefont{Kim et~al.}(2012)\citenamefont{Kim, Lamsal, Heitmann,
  Tucker, Pratt, Khan, Lee, Alam, Thaler, Ni et~al.}}]{Kim12}
\bibinfo{author}{\bibfnamefont{M.~G.} \bibnamefont{Kim}},
  \bibinfo{author}{\bibfnamefont{J.}~\bibnamefont{Lamsal}},
  \bibinfo{author}{\bibfnamefont{T.~W.} \bibnamefont{Heitmann}},
  \bibinfo{author}{\bibfnamefont{G.~S.} \bibnamefont{Tucker}},
  \bibinfo{author}{\bibfnamefont{D.~K.} \bibnamefont{Pratt}},
  \bibinfo{author}{\bibfnamefont{S.~N.} \bibnamefont{Khan}},
  \bibinfo{author}{\bibfnamefont{Y.~B.} \bibnamefont{Lee}},
  \bibinfo{author}{\bibfnamefont{A.}~\bibnamefont{Alam}},
  \bibinfo{author}{\bibfnamefont{A.}~\bibnamefont{Thaler}},
  \bibinfo{author}{\bibfnamefont{N.}~\bibnamefont{Ni}}, \bibnamefont{et~al.},
  \bibinfo{journal}{Phys. Rev. Lett.} \textbf{\bibinfo{volume}{109}},
  \bibinfo{pages}{167003} (\bibinfo{year}{2012}).

\bibitem[{\citenamefont{Abernathy et~al.}(2012)\citenamefont{Abernathy, Stone,
  Loguillo, Lucas, Delaire, Tang, Lin, and Fultz}}]{Abernathy12}
\bibinfo{author}{\bibfnamefont{D.~L.} \bibnamefont{Abernathy}},
  \bibinfo{author}{\bibfnamefont{M.~B.} \bibnamefont{Stone}},
  \bibinfo{author}{\bibfnamefont{M.~J.} \bibnamefont{Loguillo}},
  \bibinfo{author}{\bibfnamefont{M.~S.} \bibnamefont{Lucas}},
  \bibinfo{author}{\bibfnamefont{O.}~\bibnamefont{Delaire}},
  \bibinfo{author}{\bibfnamefont{X.}~\bibnamefont{Tang}},
  \bibinfo{author}{\bibfnamefont{J.~Y.~Y.} \bibnamefont{Lin}},
  \bibnamefont{and} \bibinfo{author}{\bibfnamefont{B.}~\bibnamefont{Fultz}},
  \bibinfo{journal}{Review of Scientific Instruments}
  \textbf{\bibinfo{volume}{83}}, \bibinfo{eid}{015114} (\bibinfo{year}{2012}).

\bibitem[{\citenamefont{Lester et~al.}(2009)\citenamefont{Lester, Chu,
  Analytis, Capelli, Erickson, Condron, Toney, Fisher, and Hayden}}]{Lester09}
\bibinfo{author}{\bibfnamefont{C.}~\bibnamefont{Lester}},
  \bibinfo{author}{\bibfnamefont{J.-H.} \bibnamefont{Chu}},
  \bibinfo{author}{\bibfnamefont{J.~G.} \bibnamefont{Analytis}},
  \bibinfo{author}{\bibfnamefont{S.~C.} \bibnamefont{Capelli}},
  \bibinfo{author}{\bibfnamefont{A.~S.} \bibnamefont{Erickson}},
  \bibinfo{author}{\bibfnamefont{C.~L.} \bibnamefont{Condron}},
  \bibinfo{author}{\bibfnamefont{M.~F.} \bibnamefont{Toney}},
  \bibinfo{author}{\bibfnamefont{I.~R.} \bibnamefont{Fisher}},
  \bibnamefont{and} \bibinfo{author}{\bibfnamefont{S.~M.}
  \bibnamefont{Hayden}}, \bibinfo{journal}{Phys. Rev. B}
  \textbf{\bibinfo{volume}{79}}, \bibinfo{pages}{144523}
  (\bibinfo{year}{2009}).

\bibitem[{\citenamefont{Kreyssig et~al.}(2010)\citenamefont{Kreyssig, Kim,
  Nandi, Pratt, Tian, Zarestky, Ni, Thaler, Bud'ko, Canfield
  et~al.}}]{Kreyssig10}
\bibinfo{author}{\bibfnamefont{A.}~\bibnamefont{Kreyssig}},
  \bibinfo{author}{\bibfnamefont{M.~G.} \bibnamefont{Kim}},
  \bibinfo{author}{\bibfnamefont{S.}~\bibnamefont{Nandi}},
  \bibinfo{author}{\bibfnamefont{D.~K.} \bibnamefont{Pratt}},
  \bibinfo{author}{\bibfnamefont{W.}~\bibnamefont{Tian}},
  \bibinfo{author}{\bibfnamefont{J.~L.} \bibnamefont{Zarestky}},
  \bibinfo{author}{\bibfnamefont{N.}~\bibnamefont{Ni}},
  \bibinfo{author}{\bibfnamefont{A.}~\bibnamefont{Thaler}},
  \bibinfo{author}{\bibfnamefont{S.~L.} \bibnamefont{Bud'ko}},
  \bibinfo{author}{\bibfnamefont{P.~C.} \bibnamefont{Canfield}},
  \bibnamefont{et~al.}, \bibinfo{journal}{Phys. Rev. B}
  \textbf{\bibinfo{volume}{81}}, \bibinfo{pages}{134512}
  (\bibinfo{year}{2010}).

\bibitem[{\citenamefont{Lu et~al.}(2014{\natexlab{b}})\citenamefont{Lu, Park,
  Zhang, Luo, Nevidomskyy, Si, and Dai}}]{Lu_science_2014}
\bibinfo{author}{\bibfnamefont{X.}~\bibnamefont{Lu}},
  \bibinfo{author}{\bibfnamefont{J.~T.} \bibnamefont{Park}},
  \bibinfo{author}{\bibfnamefont{R.}~\bibnamefont{Zhang}},
  \bibinfo{author}{\bibfnamefont{H.}~\bibnamefont{Luo}},
  \bibinfo{author}{\bibfnamefont{A.~H.} \bibnamefont{Nevidomskyy}},
  \bibinfo{author}{\bibfnamefont{Q.}~\bibnamefont{Si}}, \bibnamefont{and}
  \bibinfo{author}{\bibfnamefont{P.}~\bibnamefont{Dai}},
  \bibinfo{journal}{Science} \textbf{\bibinfo{volume}{345}},
  \bibinfo{pages}{6197} (\bibinfo{year}{2014}{\natexlab{b}}).

\bibitem[{\citenamefont{Pajerowski et~al.}(2013)\citenamefont{Pajerowski,
  Rotundu, Lynn, and Birgeneau}}]{Pajerowski13}
\bibinfo{author}{\bibfnamefont{D.~M.} \bibnamefont{Pajerowski}},
  \bibinfo{author}{\bibfnamefont{C.~R.} \bibnamefont{Rotundu}},
  \bibinfo{author}{\bibfnamefont{J.~W.} \bibnamefont{Lynn}}, \bibnamefont{and}
  \bibinfo{author}{\bibfnamefont{R.~J.} \bibnamefont{Birgeneau}},
  \bibinfo{journal}{Phys. Rev. B} \textbf{\bibinfo{volume}{87}},
  \bibinfo{pages}{134507} (\bibinfo{year}{2013}).

\bibitem[{\citenamefont{Huang et~al.}(2008)\citenamefont{Huang, Qiu, Bao,
  Green, Lynn, Gasparovic, Wu, Wu, and Chen}}]{Huang08}
\bibinfo{author}{\bibfnamefont{Q.}~\bibnamefont{Huang}},
  \bibinfo{author}{\bibfnamefont{Y.}~\bibnamefont{Qiu}},
  \bibinfo{author}{\bibfnamefont{W.}~\bibnamefont{Bao}},
  \bibinfo{author}{\bibfnamefont{M.~A.} \bibnamefont{Green}},
  \bibinfo{author}{\bibfnamefont{J.~W.} \bibnamefont{Lynn}},
  \bibinfo{author}{\bibfnamefont{Y.~C.} \bibnamefont{Gasparovic}},
  \bibinfo{author}{\bibfnamefont{T.}~\bibnamefont{Wu}},
  \bibinfo{author}{\bibfnamefont{G.}~\bibnamefont{Wu}}, \bibnamefont{and}
  \bibinfo{author}{\bibfnamefont{X.~H.} \bibnamefont{Chen}},
  \bibinfo{journal}{Phys. Rev. Lett.} \textbf{\bibinfo{volume}{101}},
  \bibinfo{pages}{257003} (\bibinfo{year}{2008}).

\bibitem[{\citenamefont{Lumsden et~al.}(2009)\citenamefont{Lumsden,
  Christianson, Parshall, Stone, Nagler, MacDougall, Mook, Lokshin, Egami,
  Abernathy et~al.}}]{Lumsden_2009}
\bibinfo{author}{\bibfnamefont{M.~D.} \bibnamefont{Lumsden}},
  \bibinfo{author}{\bibfnamefont{A.~D.} \bibnamefont{Christianson}},
  \bibinfo{author}{\bibfnamefont{D.}~\bibnamefont{Parshall}},
  \bibinfo{author}{\bibfnamefont{M.~B.} \bibnamefont{Stone}},
  \bibinfo{author}{\bibfnamefont{S.~E.} \bibnamefont{Nagler}},
  \bibinfo{author}{\bibfnamefont{G.~J.} \bibnamefont{MacDougall}},
  \bibinfo{author}{\bibfnamefont{H.~A.} \bibnamefont{Mook}},
  \bibinfo{author}{\bibfnamefont{K.}~\bibnamefont{Lokshin}},
  \bibinfo{author}{\bibfnamefont{T.}~\bibnamefont{Egami}},
  \bibinfo{author}{\bibfnamefont{D.~L.} \bibnamefont{Abernathy}},
  \bibnamefont{et~al.}, \bibinfo{journal}{Phys. Rev. Lett.}
  \textbf{\bibinfo{volume}{102}}, \bibinfo{pages}{107005}
  (\bibinfo{year}{2009}).

\bibitem[{\citenamefont{Harriger et~al.}(2009)\citenamefont{Harriger,
  Schneidewind, Li, Zhao, Li, Lu, Dong, Zhou, Zhao, Hu et~al.}}]{Harriger09}
\bibinfo{author}{\bibfnamefont{L.~W.} \bibnamefont{Harriger}},
  \bibinfo{author}{\bibfnamefont{A.}~\bibnamefont{Schneidewind}},
  \bibinfo{author}{\bibfnamefont{S.}~\bibnamefont{Li}},
  \bibinfo{author}{\bibfnamefont{J.}~\bibnamefont{Zhao}},
  \bibinfo{author}{\bibfnamefont{Z.}~\bibnamefont{Li}},
  \bibinfo{author}{\bibfnamefont{W.}~\bibnamefont{Lu}},
  \bibinfo{author}{\bibfnamefont{X.}~\bibnamefont{Dong}},
  \bibinfo{author}{\bibfnamefont{F.}~\bibnamefont{Zhou}},
  \bibinfo{author}{\bibfnamefont{Z.}~\bibnamefont{Zhao}},
  \bibinfo{author}{\bibfnamefont{J.}~\bibnamefont{Hu}}, \bibnamefont{et~al.},
  \bibinfo{journal}{Phys. Rev. Lett.} \textbf{\bibinfo{volume}{103}},
  \bibinfo{pages}{087005} (\bibinfo{year}{2009}).

\bibitem[{\citenamefont{Tucker et~al.}(2014)\citenamefont{Tucker, Fernandes,
  Pratt, Thaler, Ni, Marty, Christianson, Lumsden, Sales, Sefat
  et~al.}}]{Tucker14}
\bibinfo{author}{\bibfnamefont{G.~S.} \bibnamefont{Tucker}},
  \bibinfo{author}{\bibfnamefont{R.~M.} \bibnamefont{Fernandes}},
  \bibinfo{author}{\bibfnamefont{D.~K.} \bibnamefont{Pratt}},
  \bibinfo{author}{\bibfnamefont{A.}~\bibnamefont{Thaler}},
  \bibinfo{author}{\bibfnamefont{N.}~\bibnamefont{Ni}},
  \bibinfo{author}{\bibfnamefont{K.}~\bibnamefont{Marty}},
  \bibinfo{author}{\bibfnamefont{A.~D.} \bibnamefont{Christianson}},
  \bibinfo{author}{\bibfnamefont{M.~D.} \bibnamefont{Lumsden}},
  \bibinfo{author}{\bibfnamefont{B.~C.} \bibnamefont{Sales}},
  \bibinfo{author}{\bibfnamefont{A.~S.} \bibnamefont{Sefat}},
  \bibnamefont{et~al.}, \bibinfo{journal}{Phys. Rev. B}
  \textbf{\bibinfo{volume}{89}}, \bibinfo{pages}{180503}
  (\bibinfo{year}{2014}).

\bibitem[{\citenamefont{Harriger et~al.}(2011)\citenamefont{Harriger, Luo, Liu,
  Frost, Hu, Norman, and Dai}}]{Harriger11}
\bibinfo{author}{\bibfnamefont{L.~W.} \bibnamefont{Harriger}},
  \bibinfo{author}{\bibfnamefont{H.~Q.} \bibnamefont{Luo}},
  \bibinfo{author}{\bibfnamefont{M.~S.} \bibnamefont{Liu}},
  \bibinfo{author}{\bibfnamefont{C.}~\bibnamefont{Frost}},
  \bibinfo{author}{\bibfnamefont{J.~P.} \bibnamefont{Hu}},
  \bibinfo{author}{\bibfnamefont{M.~R.} \bibnamefont{Norman}},
  \bibnamefont{and} \bibinfo{author}{\bibfnamefont{P.}~\bibnamefont{Dai}},
  \bibinfo{journal}{Phys. Rev. B} \textbf{\bibinfo{volume}{84}},
  \bibinfo{pages}{054544} (\bibinfo{year}{2011}).

\bibitem[{\citenamefont{Ewings et~al.}(2008)\citenamefont{Ewings, Perring,
  Bewley, Guidi, Pitcher, Parker, Clarke, and Boothroyd}}]{Ewing08}
\bibinfo{author}{\bibfnamefont{R.~A.} \bibnamefont{Ewings}},
  \bibinfo{author}{\bibfnamefont{T.~G.} \bibnamefont{Perring}},
  \bibinfo{author}{\bibfnamefont{R.~I.} \bibnamefont{Bewley}},
  \bibinfo{author}{\bibfnamefont{T.}~\bibnamefont{Guidi}},
  \bibinfo{author}{\bibfnamefont{M.~J.} \bibnamefont{Pitcher}},
  \bibinfo{author}{\bibfnamefont{D.~R.} \bibnamefont{Parker}},
  \bibinfo{author}{\bibfnamefont{S.~J.} \bibnamefont{Clarke}},
  \bibnamefont{and} \bibinfo{author}{\bibfnamefont{A.~T.}
  \bibnamefont{Boothroyd}}, \bibinfo{journal}{Phys. Rev. B}
  \textbf{\bibinfo{volume}{78}}, \bibinfo{pages}{220501}
  (\bibinfo{year}{2008}).

\bibitem[{\citenamefont{Zhao et~al.}(2008)\citenamefont{Zhao, Yao, Li, Hong,
  Chen, Chang, Ratcliff, Lynn, Mook, Chen et~al.}}]{Zhao08}
\bibinfo{author}{\bibfnamefont{J.}~\bibnamefont{Zhao}},
  \bibinfo{author}{\bibfnamefont{D.-X.} \bibnamefont{Yao}},
  \bibinfo{author}{\bibfnamefont{S.}~\bibnamefont{Li}},
  \bibinfo{author}{\bibfnamefont{T.}~\bibnamefont{Hong}},
  \bibinfo{author}{\bibfnamefont{Y.}~\bibnamefont{Chen}},
  \bibinfo{author}{\bibfnamefont{S.}~\bibnamefont{Chang}},
  \bibinfo{author}{\bibfnamefont{W.}~\bibnamefont{Ratcliff}},
  \bibinfo{author}{\bibfnamefont{J.~W.} \bibnamefont{Lynn}},
  \bibinfo{author}{\bibfnamefont{H.~A.} \bibnamefont{Mook}},
  \bibinfo{author}{\bibfnamefont{G.~F.} \bibnamefont{Chen}},
  \bibnamefont{et~al.}, \bibinfo{journal}{Phys. Rev. Lett.}
  \textbf{\bibinfo{volume}{101}}, \bibinfo{pages}{167203}
  (\bibinfo{year}{2008}).

\bibitem[{\citenamefont{McQueeney et~al.}(2008)\citenamefont{McQueeney, Diallo,
  Antropov, Samolyuk, Broholm, Ni, Nandi, Yethiraj, Zarestky, Pulikkotil
  et~al.}}]{McQueeney08}
\bibinfo{author}{\bibfnamefont{R.~J.} \bibnamefont{McQueeney}},
  \bibinfo{author}{\bibfnamefont{S.~O.} \bibnamefont{Diallo}},
  \bibinfo{author}{\bibfnamefont{V.~P.} \bibnamefont{Antropov}},
  \bibinfo{author}{\bibfnamefont{G.~D.} \bibnamefont{Samolyuk}},
  \bibinfo{author}{\bibfnamefont{C.}~\bibnamefont{Broholm}},
  \bibinfo{author}{\bibfnamefont{N.}~\bibnamefont{Ni}},
  \bibinfo{author}{\bibfnamefont{S.}~\bibnamefont{Nandi}},
  \bibinfo{author}{\bibfnamefont{M.}~\bibnamefont{Yethiraj}},
  \bibinfo{author}{\bibfnamefont{J.~L.} \bibnamefont{Zarestky}},
  \bibinfo{author}{\bibfnamefont{J.~J.} \bibnamefont{Pulikkotil}},
  \bibnamefont{et~al.}, \bibinfo{journal}{Phys. Rev. Lett.}
  \textbf{\bibinfo{volume}{101}}, \bibinfo{pages}{227205}
  (\bibinfo{year}{2008}).

\bibitem[{\citenamefont{Matan et~al.}(2009)\citenamefont{Matan, Morinaga, Iida,
  and Sato}}]{Matan09}
\bibinfo{author}{\bibfnamefont{K.}~\bibnamefont{Matan}},
  \bibinfo{author}{\bibfnamefont{R.}~\bibnamefont{Morinaga}},
  \bibinfo{author}{\bibfnamefont{K.}~\bibnamefont{Iida}}, \bibnamefont{and}
  \bibinfo{author}{\bibfnamefont{T.~J.} \bibnamefont{Sato}},
  \bibinfo{journal}{Phys. Rev. B} \textbf{\bibinfo{volume}{79}},
  \bibinfo{pages}{054526} (\bibinfo{year}{2009}).

\bibitem[{\citenamefont{Diallo et~al.}(2009)\citenamefont{Diallo, Antropov,
  Perring, Broholm, Pulikkotil, Ni, Bud'ko, Canfield, Kreyssig, Goldman
  et~al.}}]{Diallo09}
\bibinfo{author}{\bibfnamefont{S.~O.} \bibnamefont{Diallo}},
  \bibinfo{author}{\bibfnamefont{V.~P.} \bibnamefont{Antropov}},
  \bibinfo{author}{\bibfnamefont{T.~G.} \bibnamefont{Perring}},
  \bibinfo{author}{\bibfnamefont{C.}~\bibnamefont{Broholm}},
  \bibinfo{author}{\bibfnamefont{J.~J.} \bibnamefont{Pulikkotil}},
  \bibinfo{author}{\bibfnamefont{N.}~\bibnamefont{Ni}},
  \bibinfo{author}{\bibfnamefont{S.~L.} \bibnamefont{Bud'ko}},
  \bibinfo{author}{\bibfnamefont{P.~C.} \bibnamefont{Canfield}},
  \bibinfo{author}{\bibfnamefont{A.}~\bibnamefont{Kreyssig}},
  \bibinfo{author}{\bibfnamefont{A.~I.} \bibnamefont{Goldman}},
  \bibnamefont{et~al.}, \bibinfo{journal}{Phys. Rev. Lett.}
  \textbf{\bibinfo{volume}{102}}, \bibinfo{pages}{187206}
  (\bibinfo{year}{2009}).

\bibitem[{\citenamefont{Zheludev}(2007)}]{Reslib}
\bibinfo{author}{\bibfnamefont{A.}~\bibnamefont{Zheludev}},
  \emph{\bibinfo{title}{ResLib 3.4}} (\bibinfo{publisher}{Oak Ridge National
  Laboratory}, \bibinfo{year}{2007}).

\bibitem[{Tob()}]{Tobyfit}
\bibinfo{note}{Tobyfit program available at
  \url{http://tobyfit.isis.rl.ac.uk/Main_Page.}}

\bibitem[{\citenamefont{Zhao et~al.}(2009)\citenamefont{Zhao, Adroja, Yao,
  Bewley, Li, Wang, Wu, Chen, Hu, and Dai}}]{Zhao09}
\bibinfo{author}{\bibfnamefont{J.}~\bibnamefont{Zhao}},
  \bibinfo{author}{\bibfnamefont{D.~T.} \bibnamefont{Adroja}},
  \bibinfo{author}{\bibfnamefont{D.-X.} \bibnamefont{Yao}},
  \bibinfo{author}{\bibfnamefont{R.}~\bibnamefont{Bewley}},
  \bibinfo{author}{\bibfnamefont{S.}~\bibnamefont{Li}},
  \bibinfo{author}{\bibfnamefont{X.~F.} \bibnamefont{Wang}},
  \bibinfo{author}{\bibfnamefont{G.}~\bibnamefont{Wu}},
  \bibinfo{author}{\bibfnamefont{X.~H.} \bibnamefont{Chen}},
  \bibinfo{author}{\bibfnamefont{J.}~\bibnamefont{Hu}}, \bibnamefont{and}
  \bibinfo{author}{\bibfnamefont{P.}~\bibnamefont{Dai}},
  \bibinfo{journal}{Nature Phys.} \textbf{\bibinfo{volume}{5}},
  \bibinfo{pages}{555} (\bibinfo{year}{2009}).

\bibitem[{\citenamefont{Ewings et~al.}(2011)\citenamefont{Ewings, Perring,
  Gillett, Das, Sebastian, Taylor, Guidi, and Boothroyd}}]{Ewings11}
\bibinfo{author}{\bibfnamefont{R.~A.} \bibnamefont{Ewings}},
  \bibinfo{author}{\bibfnamefont{T.~G.} \bibnamefont{Perring}},
  \bibinfo{author}{\bibfnamefont{J.}~\bibnamefont{Gillett}},
  \bibinfo{author}{\bibfnamefont{S.~D.} \bibnamefont{Das}},
  \bibinfo{author}{\bibfnamefont{S.~E.} \bibnamefont{Sebastian}},
  \bibinfo{author}{\bibfnamefont{A.~E.} \bibnamefont{Taylor}},
  \bibinfo{author}{\bibfnamefont{T.}~\bibnamefont{Guidi}}, \bibnamefont{and}
  \bibinfo{author}{\bibfnamefont{A.~T.} \bibnamefont{Boothroyd}},
  \bibinfo{journal}{Phys. Rev. B} \textbf{\bibinfo{volume}{83}},
  \bibinfo{pages}{214519} (\bibinfo{year}{2011}).

\bibitem[{\citenamefont{Lester et~al.}(2010)\citenamefont{Lester, Chu,
  Analytis, Perring, Fisher, and Hayden}}]{Lester10}
\bibinfo{author}{\bibfnamefont{C.}~\bibnamefont{Lester}},
  \bibinfo{author}{\bibfnamefont{J.-H.} \bibnamefont{Chu}},
  \bibinfo{author}{\bibfnamefont{J.~G.} \bibnamefont{Analytis}},
  \bibinfo{author}{\bibfnamefont{T.~G.} \bibnamefont{Perring}},
  \bibinfo{author}{\bibfnamefont{I.~R.} \bibnamefont{Fisher}},
  \bibnamefont{and} \bibinfo{author}{\bibfnamefont{S.~M.}
  \bibnamefont{Hayden}}, \bibinfo{journal}{Phys. Rev. B}
  \textbf{\bibinfo{volume}{81}}, \bibinfo{pages}{064505}
  (\bibinfo{year}{2010}).

\bibitem[{\citenamefont{Inosov et~al.}(2009)\citenamefont{Inosov, Park,
  Bourges, Sun, Sidis, Schneidewind, Hradil, Haug, Lin, Keimer
  et~al.}}]{Inosov09}
\bibinfo{author}{\bibfnamefont{D.~S.} \bibnamefont{Inosov}},
  \bibinfo{author}{\bibfnamefont{J.~T.} \bibnamefont{Park}},
  \bibinfo{author}{\bibfnamefont{P.}~\bibnamefont{Bourges}},
  \bibinfo{author}{\bibfnamefont{D.~L.} \bibnamefont{Sun}},
  \bibinfo{author}{\bibfnamefont{Y.}~\bibnamefont{Sidis}},
  \bibinfo{author}{\bibfnamefont{A.}~\bibnamefont{Schneidewind}},
  \bibinfo{author}{\bibfnamefont{K.}~\bibnamefont{Hradil}},
  \bibinfo{author}{\bibfnamefont{D.}~\bibnamefont{Haug}},
  \bibinfo{author}{\bibfnamefont{C.~T.} \bibnamefont{Lin}},
  \bibinfo{author}{\bibfnamefont{B.}~\bibnamefont{Keimer}},
  \bibnamefont{et~al.}, \bibinfo{journal}{Nature Phys.}
  \textbf{\bibinfo{volume}{6}}, \bibinfo{pages}{178} (\bibinfo{year}{2009}).

\bibitem[{\citenamefont{Hertz}(1976)}]{Hertz}
\bibinfo{author}{\bibfnamefont{J.~A.} \bibnamefont{Hertz}},
  \bibinfo{journal}{Phys. Rev. B} \textbf{\bibinfo{volume}{14}},
  \bibinfo{pages}{1165} (\bibinfo{year}{1976}).

\bibitem[{\citenamefont{Millis}(1993)}]{Millis}
\bibinfo{author}{\bibfnamefont{A.~J.} \bibnamefont{Millis}},
  \bibinfo{journal}{Phys. Rev. B} \textbf{\bibinfo{volume}{48}},
  \bibinfo{pages}{7183} (\bibinfo{year}{1993}).

\bibitem[{\citenamefont{Moroya and Takimoto}(1995)}]{Moriya}
\bibinfo{author}{\bibfnamefont{T.}~\bibnamefont{Moroya}} \bibnamefont{and}
  \bibinfo{author}{\bibfnamefont{T.}~\bibnamefont{Takimoto}},
  \bibinfo{journal}{J. Phys. Soc. Jpn.} \textbf{\bibinfo{volume}{64}},
  \bibinfo{pages}{960} (\bibinfo{year}{1995}).

\bibitem[{\citenamefont{Aronson et~al.}(1995)\citenamefont{Aronson, Osborn,
  Robinson, Lynn, Chau, Seaman, and Maple}}]{Aronson1995}
\bibinfo{author}{\bibfnamefont{M.~C.} \bibnamefont{Aronson}},
  \bibinfo{author}{\bibfnamefont{R.}~\bibnamefont{Osborn}},
  \bibinfo{author}{\bibfnamefont{R.~A.} \bibnamefont{Robinson}},
  \bibinfo{author}{\bibfnamefont{J.~W.} \bibnamefont{Lynn}},
  \bibinfo{author}{\bibfnamefont{R.}~\bibnamefont{Chau}},
  \bibinfo{author}{\bibfnamefont{C.~L.} \bibnamefont{Seaman}},
  \bibnamefont{and} \bibinfo{author}{\bibfnamefont{M.~B.} \bibnamefont{Maple}},
  \bibinfo{journal}{Phys. Rev. Lett.} \textbf{\bibinfo{volume}{75}},
  \bibinfo{pages}{725} (\bibinfo{year}{1995}).

\bibitem[{\citenamefont{Schr\"oder et~al.}(1998)\citenamefont{Schr\"oder,
  Aeppli, Bucher, Ramazashvili, and Coleman}}]{Schroder1998}
\bibinfo{author}{\bibfnamefont{A.}~\bibnamefont{Schr\"oder}},
  \bibinfo{author}{\bibfnamefont{G.}~\bibnamefont{Aeppli}},
  \bibinfo{author}{\bibfnamefont{E.}~\bibnamefont{Bucher}},
  \bibinfo{author}{\bibfnamefont{R.}~\bibnamefont{Ramazashvili}},
  \bibnamefont{and} \bibinfo{author}{\bibfnamefont{P.}~\bibnamefont{Coleman}},
  \bibinfo{journal}{Phys. Rev. Lett.} \textbf{\bibinfo{volume}{80}},
  \bibinfo{pages}{5623} (\bibinfo{year}{1998}).

\bibitem[{\citenamefont{Keimer et~al.}(1992)\citenamefont{Keimer, Belk,
  Birgeneau, Cassanho, Chen, Greven, Kastner, Aharony, Endoh, Erwin
  et~al.}}]{Keimer1992}
\bibinfo{author}{\bibfnamefont{B.}~\bibnamefont{Keimer}},
  \bibinfo{author}{\bibfnamefont{N.}~\bibnamefont{Belk}},
  \bibinfo{author}{\bibfnamefont{R.~J.} \bibnamefont{Birgeneau}},
  \bibinfo{author}{\bibfnamefont{A.}~\bibnamefont{Cassanho}},
  \bibinfo{author}{\bibfnamefont{C.~Y.} \bibnamefont{Chen}},
  \bibinfo{author}{\bibfnamefont{M.}~\bibnamefont{Greven}},
  \bibinfo{author}{\bibfnamefont{M.~A.} \bibnamefont{Kastner}},
  \bibinfo{author}{\bibfnamefont{A.}~\bibnamefont{Aharony}},
  \bibinfo{author}{\bibfnamefont{Y.}~\bibnamefont{Endoh}},
  \bibinfo{author}{\bibfnamefont{R.~W.} \bibnamefont{Erwin}},
  \bibnamefont{et~al.}, \bibinfo{journal}{Phys. Rev. B}
  \textbf{\bibinfo{volume}{46}}, \bibinfo{pages}{14034} (\bibinfo{year}{1992}).

\bibitem[{\citenamefont{Matsuda et~al.}(1993)\citenamefont{Matsuda, Birgeneau
  et~al.}}]{Matsuda1993}
\bibinfo{author}{\bibfnamefont{M.}~\bibnamefont{Matsuda}},
  \bibinfo{author}{\bibfnamefont{R.}~\bibnamefont{Birgeneau}},
  \bibnamefont{et~al.}, \bibinfo{journal}{Journal of the Physical Society of
  Japan} \textbf{\bibinfo{volume}{62}}, \bibinfo{pages}{1702}
  (\bibinfo{year}{1993}).

\bibitem[{\citenamefont{Varma et~al.}(1989)\citenamefont{Varma, Littlewood,
  Schmitt-Rink, Abrahams, and Ruckenstein}}]{Varma1989}
\bibinfo{author}{\bibfnamefont{C.~M.} \bibnamefont{Varma}},
  \bibinfo{author}{\bibfnamefont{P.~B.} \bibnamefont{Littlewood}},
  \bibinfo{author}{\bibfnamefont{S.}~\bibnamefont{Schmitt-Rink}},
  \bibinfo{author}{\bibfnamefont{E.}~\bibnamefont{Abrahams}}, \bibnamefont{and}
  \bibinfo{author}{\bibfnamefont{A.~E.} \bibnamefont{Ruckenstein}},
  \bibinfo{journal}{Phys. Rev. Lett.} \textbf{\bibinfo{volume}{63}},
  \bibinfo{pages}{1996} (\bibinfo{year}{1989}).

\bibitem[{\citenamefont{Boeri et~al.}(2008)\citenamefont{Boeri, Dolgov, and
  Golubov}}]{Boeri08}
\bibinfo{author}{\bibfnamefont{L.}~\bibnamefont{Boeri}},
  \bibinfo{author}{\bibfnamefont{O.~V.} \bibnamefont{Dolgov}},
  \bibnamefont{and} \bibinfo{author}{\bibfnamefont{A.~A.}
  \bibnamefont{Golubov}}, \bibinfo{journal}{Phys. Rev. Lett.}
  \textbf{\bibinfo{volume}{101}}, \bibinfo{pages}{026403}
  (\bibinfo{year}{2008}).

\bibitem[{\citenamefont{Sachdev et~al.}(1995)\citenamefont{Sachdev, Read, and
  Oppermann}}]{sachdev95}
\bibinfo{author}{\bibfnamefont{S.}~\bibnamefont{Sachdev}},
  \bibinfo{author}{\bibfnamefont{N.}~\bibnamefont{Read}}, \bibnamefont{and}
  \bibinfo{author}{\bibfnamefont{R.}~\bibnamefont{Oppermann}},
  \bibinfo{journal}{Phys. Rev. B} \textbf{\bibinfo{volume}{52}},
  \bibinfo{pages}{10286} (\bibinfo{year}{1995}).

\bibitem[{\citenamefont{Sengupta and Georges}(1995)}]{Segupta95}
\bibinfo{author}{\bibfnamefont{A.~M.} \bibnamefont{Sengupta}} \bibnamefont{and}
  \bibinfo{author}{\bibfnamefont{A.}~\bibnamefont{Georges}},
  \bibinfo{journal}{Phys. Rev. B} \textbf{\bibinfo{volume}{52}},
  \bibinfo{pages}{10295} (\bibinfo{year}{1995}).

\bibitem[{\citenamefont{Georges et~al.}(2001)\citenamefont{Georges, Parcollet,
  and Sachdev}}]{Georges01}
\bibinfo{author}{\bibfnamefont{A.}~\bibnamefont{Georges}},
  \bibinfo{author}{\bibfnamefont{O.}~\bibnamefont{Parcollet}},
  \bibnamefont{and} \bibinfo{author}{\bibfnamefont{S.}~\bibnamefont{Sachdev}},
  \bibinfo{journal}{Phys. Rev. B} \textbf{\bibinfo{volume}{63}},
  \bibinfo{pages}{134406} (\bibinfo{year}{2001}).

\bibitem[{\citenamefont{Bernal et~al.}(1995)\citenamefont{Bernal, MacLaughlin,
  Lukefahr, and Andraka}}]{Bernal1995}
\bibinfo{author}{\bibfnamefont{O.~O.} \bibnamefont{Bernal}},
  \bibinfo{author}{\bibfnamefont{D.~E.} \bibnamefont{MacLaughlin}},
  \bibinfo{author}{\bibfnamefont{H.~G.} \bibnamefont{Lukefahr}},
  \bibnamefont{and} \bibinfo{author}{\bibfnamefont{B.}~\bibnamefont{Andraka}},
  \bibinfo{journal}{Phys. Rev. Lett.} \textbf{\bibinfo{volume}{75}},
  \bibinfo{pages}{2023} (\bibinfo{year}{1995}).

\bibitem[{\citenamefont{Miranda et~al.}(1996)\citenamefont{Miranda,
  Dobrosavljevic, and Kotliar}}]{Miranda1996}
\bibinfo{author}{\bibfnamefont{E.}~\bibnamefont{Miranda}},
  \bibinfo{author}{\bibfnamefont{V.}~\bibnamefont{Dobrosavljevic}},
  \bibnamefont{and} \bibinfo{author}{\bibfnamefont{G.}~\bibnamefont{Kotliar}},
  \bibinfo{journal}{J. Phys.: Condens. Matter} \textbf{\bibinfo{volume}{8}},
  \bibinfo{pages}{9871} (\bibinfo{year}{1996}).

\bibitem[{\citenamefont{MacLaughlin et~al.}(2004)\citenamefont{MacLaughlin,
  Heffner, Bernal, Ishida, Sonier, Nieuwenhuys, Maple, and
  Stewart}}]{laughlin04}
\bibinfo{author}{\bibfnamefont{D.~E.} \bibnamefont{MacLaughlin}},
  \bibinfo{author}{\bibfnamefont{R.~H.} \bibnamefont{Heffner}},
  \bibinfo{author}{\bibfnamefont{O.~O.} \bibnamefont{Bernal}},
  \bibinfo{author}{\bibfnamefont{K.}~\bibnamefont{Ishida}},
  \bibinfo{author}{\bibfnamefont{J.~E.} \bibnamefont{Sonier}},
  \bibinfo{author}{\bibfnamefont{G.~J.} \bibnamefont{Nieuwenhuys}},
  \bibinfo{author}{\bibfnamefont{M.~B.} \bibnamefont{Maple}}, \bibnamefont{and}
  \bibinfo{author}{\bibfnamefont{G.~R.} \bibnamefont{Stewart}},
  \bibinfo{journal}{J. Phys.: Condens. Matter} \textbf{\bibinfo{volume}{16}},
  \bibinfo{pages}{S4479} (\bibinfo{year}{2004}).

\bibitem[{\citenamefont{Vavilov and Chubukov}(2011)}]{Vavilov11}
\bibinfo{author}{\bibfnamefont{M.~G.} \bibnamefont{Vavilov}} \bibnamefont{and}
  \bibinfo{author}{\bibfnamefont{A.~V.} \bibnamefont{Chubukov}},
  \bibinfo{journal}{Phys. Rev. B} \textbf{\bibinfo{volume}{84}},
  \bibinfo{pages}{214521} (\bibinfo{year}{2011}).

\bibitem[{\citenamefont{Fernandes et~al.}(2012)\citenamefont{Fernandes,
  Vavilov, and Chubukov}}]{Fernandes_12}
\bibinfo{author}{\bibfnamefont{R.~M.} \bibnamefont{Fernandes}},
  \bibinfo{author}{\bibfnamefont{M.~G.} \bibnamefont{Vavilov}},
  \bibnamefont{and} \bibinfo{author}{\bibfnamefont{A.~V.}
  \bibnamefont{Chubukov}}, \bibinfo{journal}{Phys. Rev. B}
  \textbf{\bibinfo{volume}{85}}, \bibinfo{pages}{140512}
  (\bibinfo{year}{2012}).

\end{thebibliography}

\end{document}